\documentclass[twocolumn,amsmath,amssymb,nofootinbib,10pt]{revtex4-2}
\usepackage{graphicx, float,amsmath, amsmath, amssymb, hyperref}
\usepackage[braket]{qcircuit}
\usepackage{xcolor}

\graphicspath{{pictures/}}

\begin{document}

\title{Quantum circuits for the Ising spin networks}

\author{Grzegorz Czelusta}
\author{Jakub Mielczarek}
\email{jakub.mielczarek@uj.edu.pl}
\affiliation{
Institute of Theoretical Physics, Jagiellonian University, 
{\L}ojasiewicza 11, 30-348 Cracow, Poland}

\date{\today}

\begin{abstract}
Spin network states are a powerful tool for constructing 
the $SU(2)$ gauge theories on a graph. In loop quantum gravity 
(LQG), they have yielded many promising predictions, although 
progress has been limited by the computational challenge of 
dealing with high-dimensional Hilbert spaces. To explore more 
general configurations, quantum computing methods can be applied 
by representing spin network states as quantum circuits. 
In this article, we introduce an improved method for constructing 
quantum circuits for 4-valent Ising spin networks, which 
utilizes a smaller number of qubits than previous approaches. 
This has practical implications for the implementation of 
quantum circuits. We also demonstrate the procedure with 
various examples, including the construction of a 10-node 
Ising spin network state. The key ingredient of the method 
is the variational transfer of partial states, which we 
illustrate through numerous examples. Our improved construction 
provides a promising avenue for further exploring the potential 
of quantum computing methods in quantum gravity research.
\end{abstract}

\maketitle

\section{Introduction}

Quantum circuit representation of gravitational states 
is a crucial area of research for several reasons. 
Firstly, it provides a framework for quantum simulations 
of quantum gravitational processes. Secondly, it offers 
a powerful tool to investigate the holographic properties 
of gravitational interactions. Thirdly, the quantum circuit 
representation can provide an upper bound on the quantum 
complexity of gravitational processes. Overall, the development 
of quantum circuit representations of gravitational states 
has the potential to advance our understanding of fundamental 
physics and also contribute to the development of quantum 
technologies.

In this study, we address the issue of the quantum circuit 
representation of the Ising spin network states within 
the context of loop quantum gravity (LQG) \cite{Rovelli:1997yv,Ashtekar:2004eh}. 
These states offer an intermediary level of complexity between 
the symmetry-reduced and general configurations, making 
them a valuable tool for investigating quantum collective 
phenomena in the realm of quantum gravity.

The Ising spin network states \cite{Feller:2015yta} are 
represented by graphs built out of the 4-valent nodes only. 
Furthermore, the links (holonomies) are associated with the 
fundamental $(j=1/2)$ representations of the $SU(2)$ group. 
In consequence, at the nodes, four spin-$1/2$ Hilbert spaces 
$\mathcal{H}_{\frac{1}{2}}$, associated with the holonomies,
meet. 

The invariance with respect to the local 
gauge symmetry (imposed by the Gauss law) implies that the states 
$\ket{\mathcal{I}}$ at the nodes are spanned by the invariant, 
so-called \emph{intertwiner}, spaces:
\begin{equation}
    \left|\mathcal{I}\right\rangle\in Inv_{SU\left(2\right)}\left(\mathcal{H}_{\frac{1}{2}}
    \otimes\mathcal{H}_{\frac{1}{2}}\otimes\mathcal{H}_{\frac{1}{2}}
    \otimes\mathcal{H}_{\frac{1}{2}}\right).
    \label{eq:inv}
\end{equation}
Here, the invariant Hilbert space is two-dimension, justifying 
the reason for using the Ising spin network terminology. 

The general state of a spin network can be represented as 
a product of intertwiner states at each node, i.e., 
$\otimes_n \ket{\mathcal{I}_n}$. However, it is also possible 
to consider a superposition of spin network base states, 
especially when the spin networks are constructed using 
entanglement carried by the holonomies. Such a state would 
be a superposition of different spin network states. However, 
the Gauss constraint can be used to project the state onto 
a particular spin network state. Thus, while a general spin 
network state can be represented as a product of intertwiner 
states, it can also be described as a superposition of spin 
network base states, which can be projected onto a specific 
spin network state by enforcing the Gauss constraint.

\begin{figure}
    \includegraphics[scale=0.3]{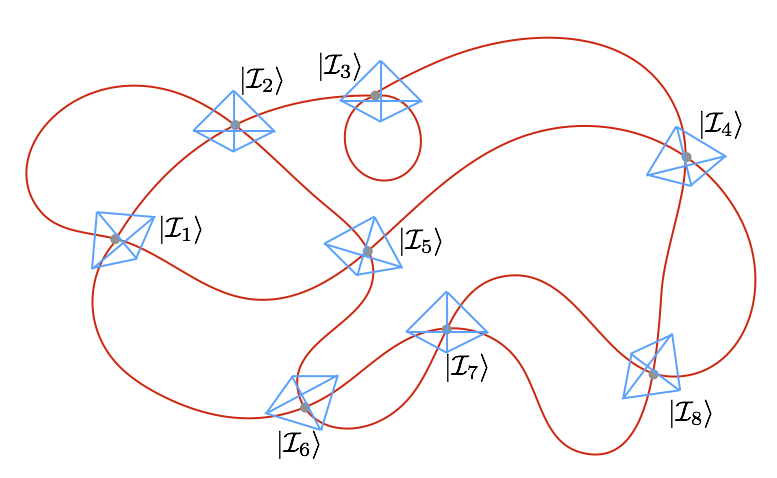}
    \caption{An exemplary Ising spin network with tetrahedra 
    being dual to the 4-valent nodes.}
    \label{fig:node}
\end{figure}

Within LQG, the geometric operators 
(e.g., area or volume) have a clear geometric interpretation 
in terms of the spin network states. Specifically, 4-valent 
nodes are associated with non-zero quanta of volume, and the 
links describe their relative adjacency. For Ising spin networks, 
the nodes are dual to tetrahedra, as depicted in Fig. \ref{fig:node}. 
The links, or holonomies, encode the adjacency of the faces of the tetrahedra.

The application of quantum computing methods to simulate 
spin networks was first explored in Ref. \cite{Li:2017gvt}. 
In this article, the Ising spin network was considered, 
allowing for the introduction of the notion of qubits. 
The model was implemented on a molecular NMR quantum 
simulator, enabling some initial quantum computations 
to be performed, specifically in the context of spin 
foam vertex amplitudes. Shortly after, another article 
on a similar subject was released \cite{Mielczarek:2018ttq}, 
which used the same concept of the ``intertwiner qubit". 
However, these studies were not conducted in the context 
of universal quantum computing, but instead focused on 
adiabatic quantum computers (quantum annealers). Overall, 
these pioneering works pave the way for future developments 
in the field.

Ref. \cite{Mielczarek:2018jsh} represents a further advancement 
in the study of quantum simulations of spin networks in LQG. 
This article describes the first-ever quantum simulations of 
nodes in the spin network using a superconducting quantum processor. 
A five-qubit superconducting quantum chip provided by IBM's cloud services 
has been used for this purpose. In addition, a method for 
evaluating transition amplitudes and the spin foam vertex 
amplitudes have been developed. Building on this work, 
Ref. \cite{Czelusta:2020ryq} introduces a quantum circuit 
that enables the preparation of a general intertwiner qubit 
state of the Ising spin network node. 

Ref. \cite{Cohen:2020jlj} proposes the use of photonic 
circuits to simulate spin foam amplitudes, which employs 
the intertwiner qubits introduced previously. This approach 
offers several advantages, including its applicability to 
spin-foam amplitudes with spin labels $j \gg 1/2$, which 
is difficult to determine using classical computations 
\cite{Dona:2019dkf,Dona:2022yyn}. Recently, Ref. \cite{vanderMeer:2022jec} 
reported the first experimental demonstration of the photonic 
approach applied to evaluating the spinfoam vertex amplitude. 

In Ref. \cite{Zhang:2020lwi}, a significant step towards 
more advanced simulations of spin foam amplitudes was 
taken using a 10-qubit superconducting quantum processor. 
This marks the most sophisticated simulation of this kind 
to date. The implementation of intertwiner qubits directly 
allowed for the achievement of results, with only a single 
logical qubit needed to encode a node of a spin network 
and five for computing a spin foam vertex amplitude 
in Ooguri's model \cite{Ooguri:1992eb}. However, the 
construction of relevant states is not straightforward 
and requires the application of optimization techniques. 
Nonetheless, the approach shows promise since it can 
potentially simulate spin networks with four times more 
nodes with the same quantum computing resources. Additionally, 
this approach has the potential to extend beyond the 
$j = 1/2$ case, making it an exciting avenue for further exploration.

In this article, we present a novel extension of existing 
methods to the case of spin networks with an arbitrary 
number of nodes and an arbitrary network structure, although 
we limit our analysis to 4-valent nodes and links with 
$1/2$ spins. Specifically, we focus on spin networks 
corresponding to \emph{vector geometries} \cite{Barrett:2009as}, 
although our method can be easily extended to other types 
of networks. Our main objective is to develop a methodology 
to represent any Ising spin network as a quantum circuit, 
to be executed on a quantum processor, and to measure 
relevant physical quantities.

This article is organized as follows. In Section 
\ref{Sec:SingleNode}, we introduce a new quantum 
circuit for the node of the Ising spin network, 
which will play a crucial role in the overall 
procedure described in this article. In Section 
\ref{Sec:Gluing}, we discuss the general procedure 
for constructing a spin network from entangled 
states of the links. Drawing on the results of 
the previous sections, we propose a new projection 
scheme in Section \ref{Sec:Projection} and illustrate 
it using the example of the dipole spin network. 
Section \ref{Sec:Variational} describes the 
variational technique used to transfer the spin 
network states onto an ansatz circuit, which is 
crucial for our optimized method of constructing 
quantum circuits for arbitrary Ising spin networks. 
The method is presented in Section \ref{Sec:GluingNetworks},
where numerous examples can also be found. 
Finally, in Section \ref{Sec:Results}, we provide 
further details of the computations, including 
the use of quantum hardware. We summarize our 
results and suggest further research directions 
in the discussed area in Section \ref{Sec:Summary}.

\section{A single node circuit}
\label{Sec:SingleNode}

In our previous article \cite{Czelusta:2020ryq}, 
we presented a quantum circuit that can create 
an arbitrary state for a single node in the 
Ising spin network. This state can be expressed 
in the following form:
\begin{equation}
\begin{split}
    |\mathcal{I}\rangle &= \cos(\theta/2) |\iota_0\rangle+e^{i\phi} \sin(\theta/2)|\iota_1\rangle\\
                        &= \frac{c_1}{\sqrt{2}} ( |0011\rangle +|1100\rangle )\\ 
                        &+ \frac{c_2}{\sqrt{2}} ( |0101\rangle +|1010\rangle )\\ 
                        &+ \frac{c_3}{\sqrt{2}} ( |0110\rangle +|1001\rangle ),\\
                        \label{QubitState}
\end{split}
\end{equation}
where $\theta$ and $\phi$ are angles on the Bloch sphere and
$\{ |\iota_0\rangle,|\iota_1\rangle \}$ form an orthonormal
basis of the invariant subspace. The coefficients $c_1$, $c_2$ 
and $c_3$ are certain functions of $\theta$ and $\phi$ 
(see Ref. \cite{Czelusta:2020ryq} for details), satisfying 
the two conditions:
\begin{align}
\sum_{i=1}^3|c_i|^2=1, \ \text{and} \ \ \sum_{i=1}^3c_i= 0.
\end{align}

\begin{figure}[h!]
	\leavevmode
	\centering
	\Qcircuit @C=1em @R=1em {
		\lstick{\ket{0}} & \gate{H}  & \qw & \qw & \qw & \qw  & \ctrl{1} & \ctrl{2} & \ctrl{3} & \qw\\
		\lstick{\ket{0}} & \gate{U} & \ctrl{1} & \ctrlo{1} & \qw & \ctrlo{2} &  \targ & \qw & \qw & \qw\\
		\lstick{\ket{0}} & \qw & \gate{V} & \targ & \ctrlo{1} & \qw &  \qw & \targ & \qw & \qw\\
		\lstick{\ket{0}} & \qw & \qw & \qw & \targ & \targ &  \qw & \qw  & \targ & \qw\\
	}
	\caption{A quantum circuit generating a state of 
	a single node of the Ising spin network, introduced in Ref. \cite{Czelusta:2020ryq}.}
	\label{fig:old_circ}
\end{figure}
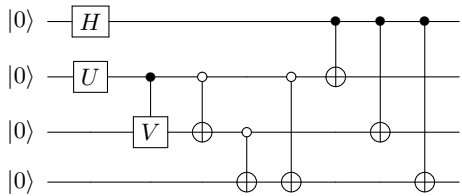

In Ref. \cite{Czelusta:2020ryq}, a quantum circuit is 
proposed to generate the state (\ref{QubitState}). 
The circuit's structure is depicted in Fig. \ref{fig:old_circ}.

Here, the gates $V$ and $U$ are given by the following unitary 
matrices:
\begin{equation}
V=\left(
\begin{array}{cc}
-\frac{c_2}{\sqrt{|c_2|^2+|c_3|^2}} & \frac{c_3^*}{\sqrt{|c_2|^2+|c_3|^2}} \\
-\frac{c_3}{\sqrt{|c_2|^2+|c_3|^2}} & -\frac{c_2^*}{\sqrt{|c_2|^2+|c_3|^2}}
\end{array}
\right),	
\end{equation}
\begin{equation}
U=\left(
\begin{array}{cc}
c_1 & \sqrt{|c_2|^2+|c_3|^2} \\
-\sqrt{|c_2|^2+|c_3|^2} & c_1^*
\end{array}	
\right).
\end{equation}

A new version of the circuit is proposed in this article, 
which involves defining the operator $\hat{W}$. This 
operator can transform a single-qubit state 
($\alpha \ket{0}+\beta \ket{1}$) into the four-qubit intertwiner state.
\begin{equation}
    \hat{W}\left(\alpha|0\rangle+\beta|1\rangle\right)|000\rangle
    =|\mathcal{I(\alpha,\beta)}\rangle
    =\alpha|\iota_0\rangle+\beta|\iota_1\rangle .
    \label{OperatorWEquation}
\end{equation}

The construction is made such that the coefficients $\alpha$ and 
$\beta$ of the single-qubit state (in the $\{ \ket{0},\ket{1} \}$ 
basis) map one-to-one onto the coefficients of the intertwiner 
qubit in the $\{ \ket{\iota_0},\ket{\iota_1}\}$ basis. Therefore, 
in order to obtain an intertwiner state represented on 4 qubits,
we need to create the corresponding 1-qubit state and subsequently 
apply the operator $\hat{W}$. The property will play an essential 
role in our further considerations. 

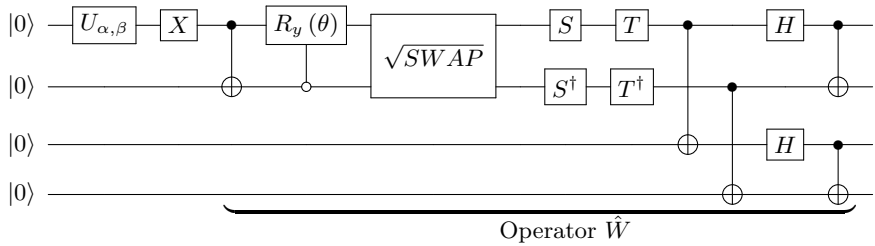
\begin{figure*}[ht!]
    \leavevmode
    \centering
    \Qcircuit @C=1em @R=1em {
        \lstick{\ket{0}} & \gate{U_{\alpha,\beta}} &\gate{X} & \ctrl{1} & \gate{R_y\left(\theta\right)} & \multigate{1}{\sqrt{SWAP}} &  \qw & \gate{S}& \gate{T} & \ctrl{2} & \qw & \gate{H} & \ctrl{1} & \qw\\
        \lstick{\ket{0}}& \qw  & \qw & \targ & \ctrlo{-1} & \ghost{\sqrt{SWAP}}& \qw & \gate{S^\dagger} & \gate{T^\dagger} & \qw & \ctrl{2} & \qw & \targ & \qw\\
        \lstick{\ket{0}} & \qw & \qw & \qw & \qw & \qw & \qw & \qw & \qw & \targ & \qw & \gate{H} & \ctrl{1} & \qw\\
        \lstick{\ket{0}}& \qw & \qw & \qw & \qw & \qw & \qw & \qw & \qw & \qw & \targ &\qw & \targ &\qw  \\
        & & & & & & & \mbox{Operator $\hat{W}$}
        \gategroup{4}{4}{4}{13}{.6em}{_)}
    }
    \caption{Quantum circuit for the operator $\hat{W}$ which turns the 
    state $\left(\alpha|0\rangle+\beta|1\rangle\right)|000\rangle$ to $|\mathcal{I(\alpha,\beta)}\rangle=\alpha|\iota_0\rangle+\beta|\iota_1\rangle$. Here, the gate $\hat{U}_{\alpha,\beta}$ acts as follows: 
    $\hat{U}\ket{0} =\alpha \ket{0}+\beta \ket{1}$. The rotation angle of the 
    rotation gate $R_y(\theta)$ is fixed to be $\theta=2\arccos{\frac{1}{\sqrt{3}}}$. The $\sqrt{SWAP}$ is a squared 
    SWAP gate, the matrix representation of which is given in Eq. \ref{squaredSWAP}. Furthermore, $S=\sqrt{Z}$ and $T=\sqrt[4]{Z}$, 
    where $Z$ denotes Pauli $Z$ gate, $X$ is the Pauli $X$ gate
    and $H$ is the Hadamard gate.}
    \label{fig:Wacts}
\end{figure*}

The quantum circuit representation of the Eq. \ref{OperatorWEquation} 
is shown in Fig. \ref{fig:Wacts}.
In the definition of the operator $\hat{W}$, one applies the rotation 
operator $R_y(\theta)$, where the rotation angle is fixed to be 
$\theta=2\arccos{\frac{1}{\sqrt{3}}}$. Furthermore, the square root 
of the $SWAP$ gate has the following matrix representation:
\begin{equation}
        \sqrt{SWAP}=\left(\begin{array}{cccc}
            1&0&0&0\\
            0&\frac{1+i}{2}&\frac{1-i}{2}&0\\
            0&\frac{1-i}{2}&\frac{1+i}{2}&0\\
            &0&0&1\\
        \end{array}\right).
        \label{squaredSWAP}
    \end{equation}
    
In the Appendix, one can find an explicit circuit for 
the $\sqrt{SWAP}$ gate. This circuit is expressed in
terms of elementary single- and two-qubit gates, 
which makes it useful for practical implementations. 
Furthermore, the Appendix provides an alternative 
circuit for the $\hat{W}$ circuit, with 
employed variational ansatz. 

\section{Gluing tetrahedra}
\label{Sec:Gluing}

Spin network states can be constructed 
by first defining the states of the links, 
and then projecting each node onto the 
intertwiner basis states. We can denote 
the state of a link $l$ as $|\alpha^{(l)}\rangle$. 
With this notation, the spin network state 
can be expressed as follows:
\begin{equation}
	\left|\Gamma,\alpha_l\right\rangle =
	\hat{P}_\Gamma\bigotimes_l\left|\alpha_l\right\rangle,
\end{equation}
where $\hat{P}_\Gamma$ is the projection operator 
(satisfying $\hat{P}_\Gamma^2=\hat{P}_\Gamma$), 
associated with the Gauss constraint, which can be 
expressed as follows:
\begin{equation}
	\hat{P}_\Gamma :=\sum_{j_l,\mathcal{I}_n}\left|\Gamma,j_l,
	\mathcal{I}_n\right\rangle
	\langle\left.\Gamma,j_l,\mathcal{I}_n\right|.
	\label{eq:projection}
\end{equation}

The classical phase-space structure that arises from 
spin networks is called twisted geometry \cite{Freidel:2010aq}. A typical 
configuration of this geometry corresponds to a collection 
of uncorrelated tetrahedra (or, in the case of networks 
with higher valency, polyhedra). This uncorrelated structure 
of the twisted geometry is reflected by the uncorrelated 
structure of the spin network basis states:
\begin{equation}
    \left|\Gamma,j_l,\mathcal{I}_n\right\rangle = \bigotimes_n|\mathcal{I}_n\rangle,
\end{equation}
which is a product state, not exhibiting entanglement
between the nodes. 

In this article, we will focus on vector geometries, 
which have a more rigid structure compared to other 
geometries. Specifically, in vector geometries, the 
normals to adjacent faces in neighboring tetrahedra 
are oriented back-to-back. Despite this specific 
focus, it is important to note that the method presented 
in this article is completely general and applicable 
to arbitrary types of links.

It has been justified in Ref. \cite{Baytas:2018wjd}
that the vector geometry can be obtained by using 
the squeezed states on the links:
\begin{equation}
\left|\mathcal{B},\lambda_l\right\rangle = \left(1-\left|\lambda_l\right|^2\right)
\sum_j\sqrt{2j+1}\lambda_l^{2j}
\left|\mathcal{B},j\right\rangle,
\end{equation}
where $\lambda_l \in \mathbb{C}$ is a free parameter 
of the state. The $\left|\mathcal{B},j\right\rangle$ 
is a singlet state of spin $j$, which is maximally entangled
\begin{equation}
	\left|\mathcal{B},j\right\rangle=\frac{1}{\sqrt{2j+1}}\sum_{m=-j}^j\left(-1\right)^{j-m}\left|j,m\right\rangle_s\left|j,-m\right\rangle_t.
\end{equation}
Subsequently, projection on a spin-network basis states is 
performed, leading to:
\begin{equation}
	\left|\Gamma,\mathcal{B},\lambda_l\right\rangle = \hat{P}_\Gamma\bigotimes_l\left|\mathcal{B},\lambda_l\right\rangle.
\end{equation}
Specifically, for the case of the Ising spin networks ($j=1/2$) 
we obtain:
\begin{equation}
	\left|\mathcal{B},j\right\rangle=\frac{1}{\sqrt{2}}\left(\left|\frac{1}{2},\frac{1}{2}\right\rangle\left|\frac{1}{2},-\frac{1}{2}\right\rangle-\left|\frac{1}{2},-\frac{1}{2}\right\rangle\left|\frac{1}{2},\frac{1}{2}\right\rangle\right)
\end{equation}
or, equivalently, in the qubit notation:
\begin{equation}
	\left|\mathcal{B},\frac{1}{2}\right\rangle
	=\frac{1}{\sqrt{2}}\left(\left|01\right\rangle
	-\left|10\right\rangle\right).
	\label{eq:bell}
\end{equation}

Previous studies have focused on spin networks 
constructed from the singlet state \ref{eq:bell}. 
In particular, Refs. \cite{Li:2017gvt,Czelusta:2020ryq} 
conducted quantum simulations of simple spin 
networks (dipole and pentagram) using this singlet 
state. Notably, it has been observed that the states 
of these spin networks correspond to the PEPS 
tensor networks \cite{Czelusta:2020ryq} implying the 
area law for entanglement entropy \cite{Han:2016xmb,Feller:2017jqx,Bianchi:2018fmq}. 

One potential drawback of this approach is that 
the number of qubits involved in the computation 
can quickly become quite large. Specifically, 
since each link in the Ising spin network state 
corresponds to a 2-qubit state, a network with 
$l$ links requires $2l$ qubits to be initially 
involved. For example, the pentagram spin network 
has $l=10$, which means that $20$ qubits are 
required initially. While applying the Gauss 
projection at the nodes can reduce the Hilbert 
space to just $5$ intertwiner qubits, this still 
means that the majority of the initial quantum 
resources are not used by the final state. This 
inefficiency becomes even more pronounced when 
dealing with more complicated spin networks.

To address this issue, a method for constructing 
Ising spin networks that uses far fewer quantum 
resources is needed. In the subsequent sections, 
we propose a concrete procedure for achieving this goal.

\section{Projection onto the intertwiner subspace}
\label{Sec:Projection}

This section outlines the initial stage of constructing 
Ising spin networks, which involves reducing the number 
of qubits required for the process.

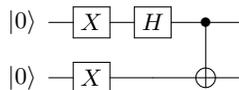
\begin{figure}[ht!]
        \leavevmode
        \centering
        \Qcircuit @C=1em @R=1em {
            \lstick{\ket{0}} &  \gate{X} & \gate{H} & \ctrl{1} & \qw\\
            \lstick{\ket{0}} &  \gate{X} & \qw & \targ & \qw \\
        }
        \caption{Quantum circuit for the Bell state (\ref{eq:bell}) .}
        \label{fig:bell_circ}
\end{figure}

For this purpose, let us consider the states 
of the links, in accordance with the discussion 
presented in the previous section. For the Bell 
states (\ref{eq:bell}) of the links, the corresponding
quantum circuit, employing elementary gates, is  
shown in Fig. \ref{fig:bell_circ}.

In our previous studies in Refs. 
\cite{Mielczarek:2018jsh,Czelusta:2020ryq}, the 
states of the links were projected onto the 
spin network basis states, generated by the sequence 
of the quantum circuits shown in Fig. \ref{fig:old_circ}.
In this way the quantum amplitudes of the entangled 
links in the spin network basis were reconstructed. 
However, the approach does not lead to the final 
state directly. 

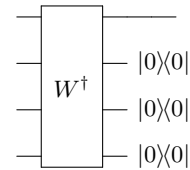
\begin{figure}[ht!]
        \leavevmode
        \centering
        \Qcircuit @C=1em @R=1em {
            & \multigate{3}{W^\dagger} & \qw & \qw & \qw\\
            & \ghost{W^\dagger} & \qw & \ket{0} & \bra{0}\\
            & \ghost{W^\dagger}& \qw & \ket{0} & \bra{0}\\
            & \ghost{W^\dagger}& \qw & \ket{0} & \bra{0}\\
        }
        \caption{Projection operator on the 
        intertwiner subspace, expressed in one-qubit representation.}
        \label{fig:projection_circ}
    \end{figure}

Here, we observe that the $\hat{W}$ operator, 
introduced in Fig. \ref{fig:Wacts}, can be used 
to define projection operator, equipped with 
the mapping onto a single-qubit state. For this 
purpose, we apply $\hat{W}^\dagger$ and then 
project some of the qubits (see Fig. \ref{fig:projection_circ}) 
onto the $\left|0\right\rangle$ states.

To achieve the desired projection, measurements 
are conducted, and only those returning $|0\rangle$ 
for the last three qubits are accepted. In the 
chosen outcomes, the first qubit stores the 
intertwiner state. 
    
Although vector geometries are utilized in this work, 
the proposed method can be adapted to other types of 
spin networks. To achieve this, alternative states, 
rather than Bell pairs, must be prepared and subsequently 
projected. Notably, the operator $\hat{W}$ described 
in this study is specific to Ising spin networks. 
Nevertheless, its extension to higher dimensions of
the intertwiner space is likely feasible.

\subsection{Dipole}

\begin{figure}[ht!]
        \includegraphics[scale=0.25]{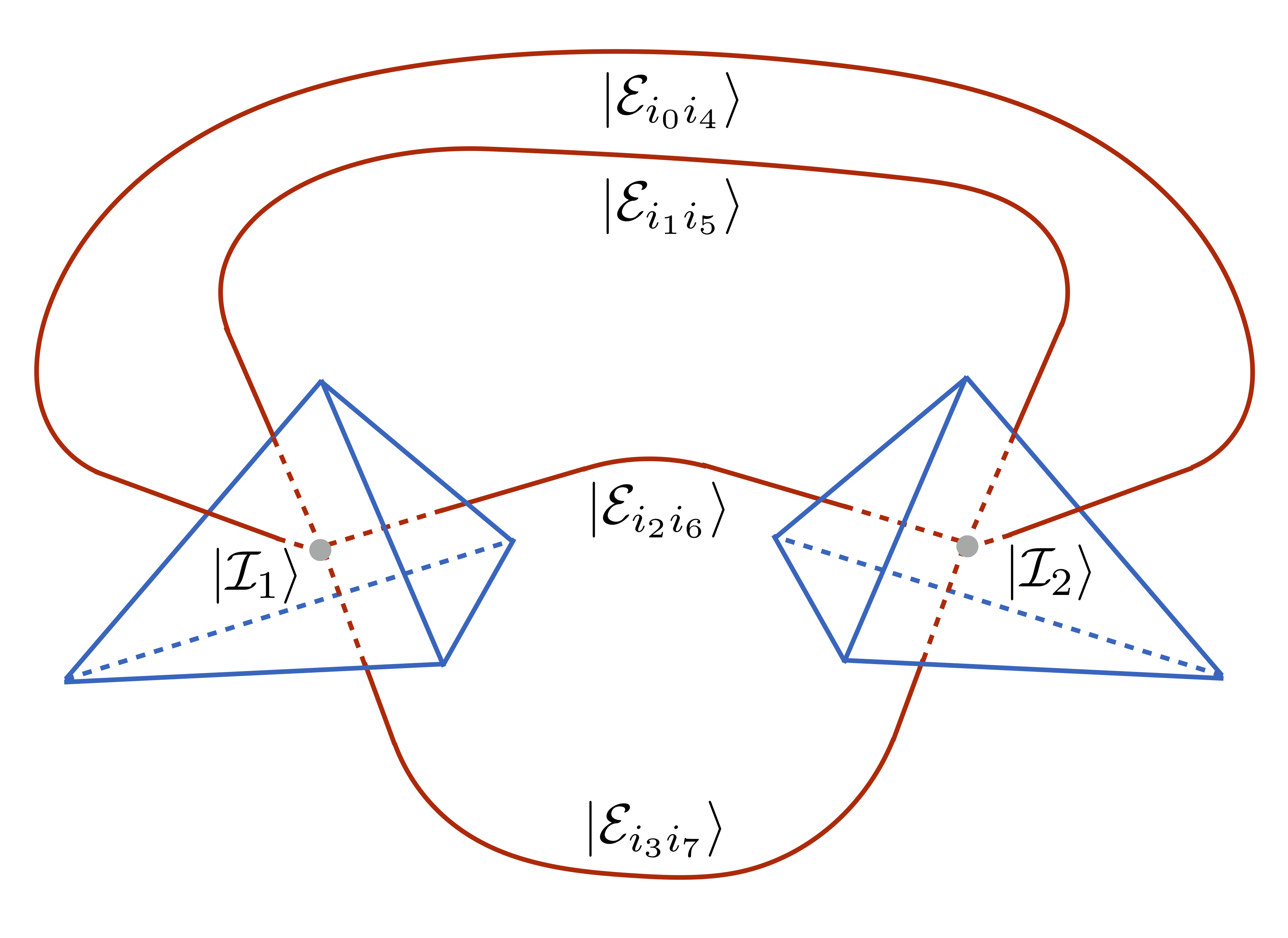}
        \caption{The dipole spin network.}
        \label{fig:dipole}
\end{figure}

To demonstrate the newly introduced projection 
technique, we analyze its application to the 
dipole spin network, which is presented in 
Fig. \ref{fig:dipole}. 

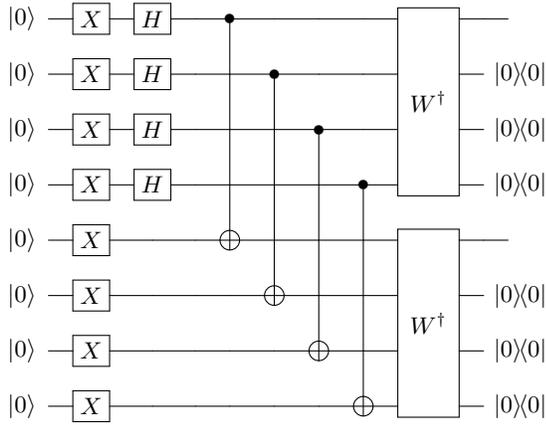
\begin{figure}[ht!]
        \leavevmode
        \centering
        \Qcircuit @C=1em @R=1em {
            \lstick{\ket{0}} & \gate{X} & \gate{H} & \qw & \ctrl{4} & \qw  & \qw  & \qw & \multigate{3}{W^\dagger}&\qw&\qw\\
            \lstick{\ket{0}} & \gate{X} & \gate{H} & \qw  & \qw  & \ctrl{4} & \qw  & \qw & \ghost{W^\dagger}&\qw&\ket{0}&\bra{0}\\
            \lstick{\ket{0}} & \gate{X} & \gate{H} & \qw  & \qw  & \qw  & \ctrl{4} & \qw & \ghost{W^\dagger}&\qw&\ket{0}&\bra{0}\\
            \lstick{\ket{0}} & \gate{X} & \gate{H} & \qw  & \qw  & \qw  & \qw  & \ctrl{4} & \ghost{W^\dagger}&\qw&\ket{0}&\bra{0}\\
            \lstick{\ket{0}} & \gate{X} & \qw & \qw  & \targ & \qw  & \qw  & \qw & \multigate{3}{W^\dagger}&\qw&\qw\\
            \lstick{\ket{0}} & \gate{X} & \qw & \qw  & \qw   & \targ & \qw & \qw & \ghost{W^\dagger} &\qw&\ket{0}&\bra{0}\\
            \lstick{\ket{0}} & \gate{X} & \qw & \qw  & \qw  & \qw & \targ & \qw & \ghost{W^\dagger} &\qw&\ket{0}&\bra{0}\\
            \lstick{\ket{0}} & \gate{X} & \qw & \qw  & \qw  & \qw  & \qw  & \targ & \ghost{W^\dagger} &\qw&\ket{0}&\bra{0}\\
        }
        \caption{Quantum circuit representing projection of state $\bigotimes_{l\in\Gamma_2}\left|\mathcal{B},\frac{1}{2}\right\rangle$.}
        \label{DipolCircuit}
    \end{figure}

As shown in Fig. \ref{DipolCircuit}, using the maximally 
entangled pairs $\left|\mathcal{B},\frac{1}{2}\right\rangle$ 
(generated by the circuit shown in Fig. \ref{fig:bell_circ}), 
the states of the links are created first. Then the projection 
operator (generated by the circuit shown in Fig. 
\ref{fig:projection_circ}) is applied for the two nodes. 
The unmeasured qubits (0th and 4th from the top) carry 
the state of the dipole spin network. 

By conducting quantum tomography on the 0th and 4th 
qubits and discarding any measurement results with 
non-zero values on qubits 1, 2, 3, 5, 6, and 7, 
we can extract the dipole state. This method enables 
us to obtain the density matrix of the dipole state, 
which comprises only two logical qubits, as anticipated 
for the two intertwiner qubits of the Ising dipole 
spin network.

On the other hand, the dipole state can easily be 
computed by the following contraction of the 
Wigner $4j$-symbols $\iota_{k}^{m_1m_2m_3m_4}$:
\begin{equation}
	\begin{split}
	&\left|\Gamma_2,\mathcal{B},\frac{1}{2}\right\rangle=\hat{P}_\Gamma\bigotimes_{l\in\Gamma_2}\left|\mathcal{B},\frac{1}{2}\right\rangle\\&=\sum_{k,l}\iota_{(k)}^{m_1m_2m_3m_4}\iota_{(l)m_1m_2m_3m_4}\left|\iota_k\iota_l\right\rangle\\&=\frac{1}{\sqrt{2}}\left(\left|\iota_0\iota_0\right\rangle+\left|\iota_1\iota_1\right\rangle\right).\\
	\end{split}
	\label{eq:dipole}
\end{equation}
Furthermore, the $4j$-symbols can be expressed in terms 
of the $3j$-symbols:
\begin{equation}
\begin{split}
    \iota_{k}^{m_1m_2m_3m_4} =& \sqrt{2k+1}\left(\begin{array}{ccc}
         m_1& m_2 & m \\
         j_1 & j_2 & k \\
    \end{array}\right)\cdot\\
    &\cdot g_{mm'}
    \left(\begin{array}{ccc}
         m'& m_3 & m_4 \\
         k & j_3 & j_4 \\
    \end{array}\right)\\
\end{split}
\end{equation}
where
\begin{equation}
    g_{mm'}=\delta_{m,-m'}(-1)^{j-m}.
\end{equation}
The indices of Wigner $4j$-symbols can be lowered 
and raised using the $g_{mm'}$ tensor.

As observed, the dipole spin network's final state 
can be encoded using only two of the original eight 
qubits utilized in its construction. Although the 
final 2-qubit state can be reconstructed using quantum 
tomography, it is essential to find its reduced circuit 
representation afterward. To avoid the need for quantum 
tomography, we present a variational method in the 
following section to construct the circuit.

\section{Variational transferring of quantum state}
\label{Sec:Variational}

Quantum circuit for some unknown state can be 
approximated (or even given exactly) by employing 
a quantum circuit ansatz. In particular, the 
projected state of the spin network can be transferred 
into an ansatz circuit, without knowing the explicit 
form of the state, by fixing the parameters of the 
circuit. The obtained circuit is characterised by 
smaller number of qubits that the original (unprojected) 
state, which is convenient for the quantum computing purposes. 

\begin{figure}[ht!]
        \leavevmode
        \centering
        \Qcircuit @C=1em @R=1em {
            & \gate{R_y\left(\theta_0\right)} & \ctrl{1} & \gate{R_y\left(\theta_5\right)} &\qw&\qw&\qw\\
            & \gate{R_y\left(\theta_1\right)} & \ctrl{-1} & \gate{R_y\left(\theta_6\right)} &\ctrl{1}& \gate{R_y\left(\theta_9\right)}&\qw\\
            & \gate{R_y\left(\theta_2\right)} & \ctrl{1} & \gate{R_y\left(\theta_7\right)} &\ctrl{-1}& \gate{R_y\left(\theta_{10}\right)}&\qw\\
            & \gate{R_y\left(\theta_3\right)} & \ctrl{-1} & \gate{R_y\left(\theta_8\right)} &\ctrl{1}& \gate{R_y\left(\theta_{11}\right)}&\qw\\
            & \gate{R_y\left(\theta_4\right)} & \qw & \qw&\ctrl{-1}& \gate{R_y\left(\theta_{12}\right)}&\qw\\
            & \mbox{Initial layer} & & & \mbox{1st layer}
            \gategroup{1}{2}{5}{2}{.8em}{_)}
            \gategroup{1}{3}{5}{6}{.8em}{_)}
        }
        \caption{Quantum circuit for the simplified-two-design 
        ansatz with one layer.}
    \label{fig:s2d}
    \end{figure}
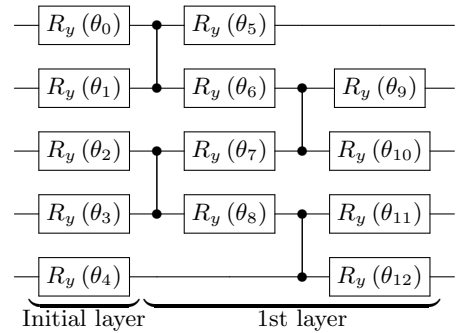

Here, we use the so-called Simplified-Two-Design ansatz, 
which consists of layers of Pauli-Y rotations and controlled-Z 
entanglers proposed in Ref. \cite{cerezo2021cost}.
The quantum circuit for the ansatz is shown in Fig. \ref{fig:s2d},
for the case with 5 quibts. 

In what follows, fidelities are used to quantify the 
quantum states generated by the ansatz circuits. Both the
classical fidelity:
\begin{equation}
    F\left(p,q\right) := \sum_i\sqrt{p_iq_i},
\end{equation}
where $p$ and $q$ are probabilities of basis states,
and the quantum fidelity:
\begin{equation}
    F(\hat{\rho}_1, \hat{\rho}_2) := \left(\text{tr}\sqrt{\sqrt{\hat{\rho}_1}\hat{\rho}_2\sqrt{\hat{\rho}_1}}\right)^2,
\end{equation}
where $\hat{\rho}_1$ and $\hat{\rho}_2$ are density 
matrices of two states will be used. 
  
\subsection{Dipole}

\begin{figure*}[ht!]
        \leavevmode
        \centering
        \Qcircuit @C=1em @R=1em {
            \lstick{\ket{0}} & \gate{X} & \gate{H} & \qw  &\ctrl{4} & \qw& \qw   & \qw  & \multigate{3}{W^\dagger}&\qw&\qw&\qw&\gate{R_y\left(\theta_0\right)}&\ctrl{4}&\gate{R_y\left(\theta_2\right)}& \qw \\
            \lstick{\ket{0}} & \gate{X} & \gate{H} & \qw  & \qw  & \ctrl{4} & \qw  & \qw & \ghost{W^\dagger}& \qw  &\ket{0}&\bra{0} \\
            \lstick{\ket{0}} & \gate{X} & \gate{H} & \qw  & \qw  & \qw  & \ctrl{4} & \qw & \ghost{W^\dagger}&\qw&\ket{0}&\bra{0}\\
            \lstick{\ket{0}} & \gate{X} & \gate{H} & \qw  & \qw  & \qw  & \qw  & \ctrl{4} & \ghost{W^\dagger}&\qw&\ket{0}&\bra{0}\\
            \lstick{\ket{0}} & \gate{X} & \qw & \qw  & \targ & \qw  & \qw  & \qw & \multigate{3}{W^\dagger}&\qw&\qw&\qw&\gate{R_y\left(\theta_1\right)}&\ctrl{-4}&\gate{R_y\left(\theta_3\right)}& \qw \\
            \lstick{\ket{0}} & \gate{X} & \qw & \qw  & \qw   & \targ & \qw  & \qw & \ghost{W^\dagger} &\qw&\ket{0}&\bra{0}\\
            \lstick{\ket{0}} & \gate{X} & \qw & \qw  & \qw & \qw  & \targ  & \qw & \ghost{W^\dagger} &\qw&\ket{0}&\bra{0}\\
            \lstick{\ket{0}} & \gate{X} & \qw & \qw  & \qw  & \qw  & \qw  & \targ & \ghost{W^\dagger} &\qw&\ket{0}&\bra{0}\\
        }
        \caption{Projection of state $\bigotimes_{l\in\Gamma_2}\left|\mathcal{B},\frac{1}{2}\right\rangle$ and transferring on simplified-two-design two-qubits ansatz.}
        \label{fig:dipole_transfer}
    \end{figure*}
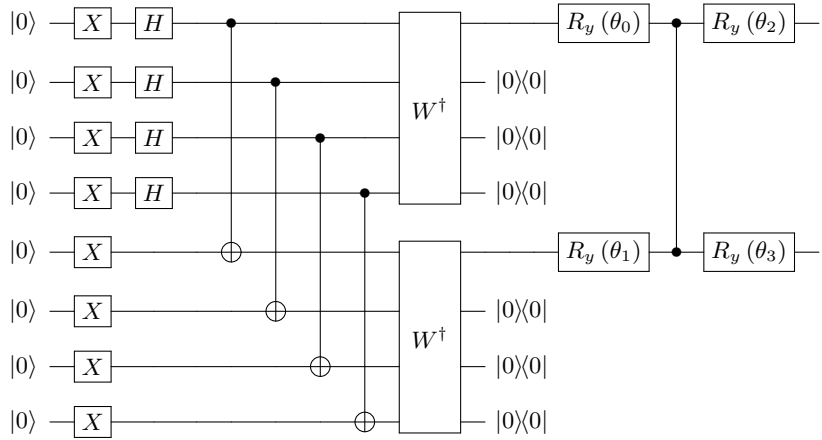

Applying the quantum circuit shown in Fig. \ref{DipolCircuit},
the resulting 0th and 4th qubit state can be projected onto 
the ansatz circuit shown in Fig. \ref{fig:s2d}. This results 
in the quantum circuit presented in Fig. \ref{fig:dipole_transfer},
which allows transferring the dipole state. 

The transfer is performed variationally  by minimizing the 
following cost function:
\begin{equation}
		\text{cost}\left(\vec{\theta}\right) = 1-\text{Prob}(q_0q_4=00),
\end{equation}    
where the parameter vector $\vec{\theta}=(\theta_1,\theta_2,
\theta_3,\theta_4)$, contains the four angles to be fixed.
We use classical optimizers, gradient descent optimizers, 
and Adam optimizers. Technical details and parameters can be 
found in our GitHub repository \cite{Repo}. 

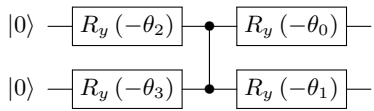
\begin{figure}[ht!]
    \leavevmode
    \centering
    \Qcircuit @C=1em @R=1em {
        \lstick{\ket{0}} &\gate{R_y\left(-\theta_2\right)}&\ctrl{1}&\gate{R_y\left(-\theta_0\right)}&\qw\\
        \lstick{\ket{0}} &\gate{R_y\left(-\theta_3\right)}&\ctrl{-1}&\gate{R_y\left(-\theta_1\right)}&\qw\\
    }
    \caption{Transferred projection of the state $\bigotimes_{l\in\Gamma_2}\left|\mathcal{B},\frac{1}{2}\right\rangle$, i.e. adjoint ansatz.}
    \label{fig:transfered_dipol}
\end{figure}

By taking the conjugation of the ansatz with the 
parameters obtained from minimizing the cost 
function, we can prepare the dipole state using 
the quantum circuit depicted in Fig.\ref{fig:transfered_dipol}. 
This circuit is designed to transform the 
initial state into the desired dipole state, 
utilizing the parameters obtained from 
the optimization process.

The transferred dipole circuit (Fig. \ref{fig:transfered_dipol}) 
has been thereafter executed on the superconducting Manila IBM 
quantum processor, resulting in the fidelities shown in Tab.
\ref{tab:fid_dipole}.

\begin{table}[h!]
    \centering
    \begin{tabular}{c|c|c}
        & without correction & with correction \\
        \hline
         classical & 0.96 & 0.9968\\
         quantum & 0.89&  0.99\\
    \end{tabular}
    \caption{Fidelities for dipole spin network.}
    \label{tab:fid_dipole}
\end{table}

In our work, we utilized measurement error mitigation 
as a means of correction. For the dipole state, 
the proposed ansatz can accurately express the state, 
leading to a fidelity that can be made arbitrarily 
close to 1. To delve into the technical details and 
results of our research, as well as the opportunity 
to experiment with our simulations, please refer 
to our GitHub repository \cite{Repo}.

The table with the found parameters of the ansatz 
for the dipole is presented in Tab. \ref{tab:dipole_params}.
\begin{table}[h!]
    \centering
    \begin{tabular}{|c|c|}
        \hline
        -1.06 & 3.65 \\
        \hline
        4.72 & 1.57 \\
        \hline
    \end{tabular}
    \caption{Determined parameters for dipole ansatz. The structure of the table corresponds to the structure of ansatz Fig. \ref{fig:s2d}, with one layer.}
    \label{tab:dipole_params}
\end{table}

\subsection{Pentagram}

In the case of the pentagram spin network shown in 
Fig. \ref{fig:pentagram}
\begin{figure}
	\includegraphics[scale=0.7]{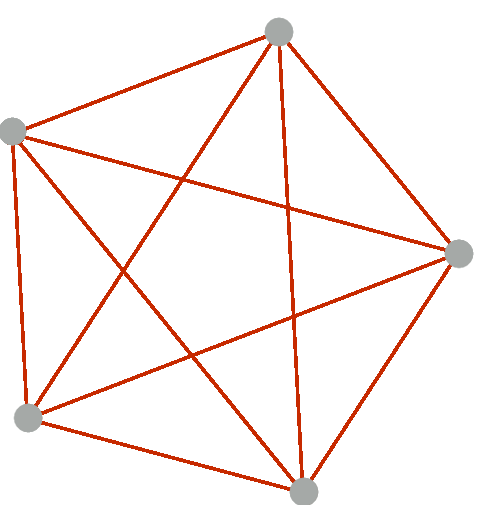}
	\caption{The pentagram spin network.}
	\label{fig:pentagram}
\end{figure}
the projected state has the form:
\begin{equation}
    \hat{P}_{\Gamma}\left|\psi\right\rangle=\sum_{\iota_{k_i}}
    	\overline{\{15j\}}\left|\iota_{k_1}\iota_{k_2}\iota_{k_3}\iota_{k_4}\iota_{k_5}\right\rangle,
\end{equation}
where the $15j$ symbol can be expressed in terms of the 
$4j$ symbols as follows:
\begin{equation}
	\begin{split}
	    \{15j\}=&\iota_1^{m_{12}m_{13}m_{14}m_{15}}\iota_{2;m_{12}}^{\qquad m_{13}m_{14}m_{15}}\iota_{3;m_{12}m_{13}}^{\qquad\quad m_{14}m_{15}}\\
	    &\iota_{4;m_{12}m_{13}m_{14}}^{\qquad\qquad\quad m_{15}}\iota_{5;m_{12}m_{13}m_{14}m_{15}}.
	\end{split}
\end{equation}

\begin{figure}[ht!]
		\leavevmode
		\centering
		\Qcircuit @C=1em @R=1em {
			\lstick{\ket{0}} & \multigate{19}{U_\psi} & \multigate{3}{W^\dagger}&\qw&\qw\\
			\lstick{\ket{0}} & \ghost{U_\psi} & \ghost{W^\dagger}&\qw&\ket{0}&\bra{0}\\
			\lstick{\ket{0}} & \ghost{U_\psi} & \ghost{W^\dagger}&\qw&\ket{0}&\bra{0}\\
			\lstick{\ket{0}} & \ghost{U_\psi} & \ghost{W^\dagger}&\qw&\ket{0}&\bra{0}\\
			\lstick{\ket{0}} & \ghost{U_\psi} & \multigate{3}{W^\dagger}&\qw&\qw&\\
			\lstick{\ket{0}} & \ghost{U_\psi} & \ghost{W^\dagger} &\qw&\ket{0}&\bra{0}\\
			\lstick{\ket{0}} & \ghost{U_\psi} & \ghost{W^\dagger} &\qw&\ket{0}&\bra{0}\\
			\lstick{\ket{0}} & \ghost{U_\psi} & \ghost{W^\dagger} &\qw&\ket{0}&\bra{0}\\
			\lstick{\ket{0}} & \ghost{U_\psi} & \multigate{3}{W^\dagger}&\qw&\qw&\\
			\lstick{\ket{0}} & \ghost{U_\psi} & \ghost{W^\dagger} &\qw&\ket{0}&\bra{0}\\
			\lstick{\ket{0}} & \ghost{U_\psi} & \ghost{W^\dagger} &\qw&\ket{0}&\bra{0}\\
			\lstick{\ket{0}} & \ghost{U_\psi} & \ghost{W^\dagger} &\qw&\ket{0}&\bra{0}\\
			\lstick{\ket{0}} & \ghost{U_\psi} & \multigate{3}{W^\dagger}&\qw&\qw&\\
			\lstick{\ket{0}} & \ghost{U_\psi} & \ghost{W^\dagger} &\qw&\ket{0}&\bra{0}\\
			\lstick{\ket{0}} & \ghost{U_\psi} & \ghost{W^\dagger} &\qw&\ket{0}&\bra{0}\\
			\lstick{\ket{0}} & \ghost{U_\psi} & \ghost{W^\dagger} &\qw&\ket{0}&\bra{0}\\
			\lstick{\ket{0}} & \ghost{U_\psi} & \multigate{3}{W^\dagger}&\qw&\qw&\\
			\lstick{\ket{0}} & \ghost{U_\psi} & \ghost{W^\dagger} &\qw&\ket{0}&\bra{0}\\
			\lstick{\ket{0}} & \ghost{U_\psi} & \ghost{W^\dagger} &\qw&\ket{0}&\bra{0}\\
			\lstick{\ket{0}} & \ghost{U_\psi} & \ghost{W^\dagger} &\qw&\ket{0}&\bra{0}\\
		}
		\caption{Quantum circuit representing projection of the state $|\psi\rangle=\bigotimes_{l\in\Gamma_5}\left|\mathcal{B},\frac{1}{2}\right\rangle$.}
		\label{fig:qcirc_pentagram}
\end{figure}
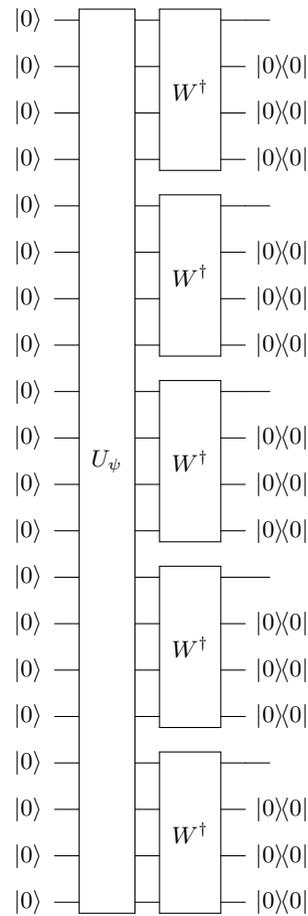

For the Ising spin network case, the state can be obtained 
using the quantum circuits presented in Fig. \ref{fig:qcirc_pentagram}.

Here, the state $|\psi\rangle=\hat{U}_\psi|0\rangle^{\otimes 20}=\bigotimes_{l\in\Gamma_5}\left|\mathcal{B},
\frac{1}{2}\right\rangle$ is a product of the singlet pairs, i.e.
multiple time use of the circuit shown in Fig. \ref{fig:bell_circ}. 
Employing similar techniques as in the case of a dipole, we can 
transfer the pentagram state on the ansatz of the type shown 
in Fig. \ref{fig:s2d}, with 3 layers.

Ansatz for the pentagram state with four layers 
and five qubits is shown in Fig. \ref{fig:s2d}. The 
explicit form of the ansatz and the found parameters 
can be found at the GitHub repository \cite{Repo}.

\begin{figure}[h!]
			\includegraphics[scale=0.13,angle=270]{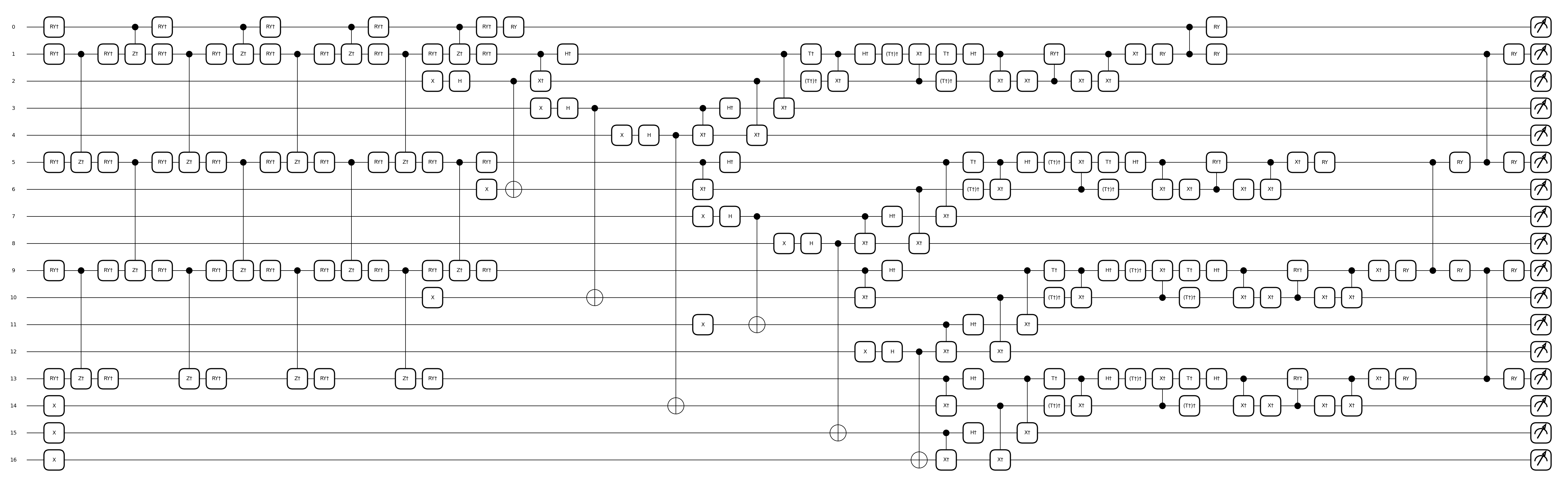}
			\caption{Quantum circuit used for the variational transfer 
			of the pentagram Ising-type spin network state (see from 
			the left side).}
			\label{fig:PentargramFullCircuit}
\end{figure}

In order to emphasize the complexity of the quantum 
circuit for the variational transfer of the pentagram 
Ising spin network state, we show its explicit form in Fig. \ref{fig:PentargramFullCircuit}.

The obtained ansatz has been executed on the IBM quantum 
computer Manila and the measured fidelities are shown in 
Tab. \ref{tab:fid_pentagram}.

\begin{table}[h!]
    \centering
    \begin{tabular}{c|c|c}
        & without correction & with correction \\
        \hline
         classical & 0.87 & 0.93\\
         quantum & 0.6 & 0.77\\
    \end{tabular}
    \caption{Fidelities for the pentagram spin network.}
    \label{tab:fid_pentagram}
\end{table}

The determined parameters of the ansatz for the 
pentagram are shown in Tab. \ref{tab:pentagram_params}.

\begin{table}[h!]
    \centering
    \begin{tabular}{|c|c|c|c|c|c|c|}
        \hline
        3.51 & 1.88 &  &  1.21 &  & 2.29 &  \\
        \hline
        3.80 & 0.77 & 1.09 & -0.15 & 4.99 & 4.96 & 4.78 \\
        \hline
        0.33 & 2.35 & 1.27 &  2.69 & 1.73 & 0.82 & 1.08 \\
        \hline
        5.04 & 0.02 & 4.82 &  4.90 & 4.65 & 6.10 & 1.96 \\
        \hline
        3.43 &  & 3.33 &   & 4.26 &  & 5.48 \\
        \hline
    \end{tabular}
    \caption{Found parameters for pentagram ansatz. The structure of the table corresponds to the structure of ansatz Fig. \ref{fig:s2d}, with three layers.}
    \label{tab:pentagram_params}
\end{table}

\section{Gluing networks}
\label{Sec:GluingNetworks}

Generating the Ising spin networks in the way presented so far 
requires the number of qubits to be four times greater than 
the number of nodes. This is becoming problematic while attempting 
to simulate higher valence spin networks, which are 
relevant e.g. to study collective properties in quantum gravity. 
However, as we have shown in the previous section, states of the 
Ising spin networks can eventually be transferred to the quantum 
circuit employing the number of logical qubits equal to the 
number of nodes. 

\begin{figure}[ht!]
			\includegraphics[scale=0.8]{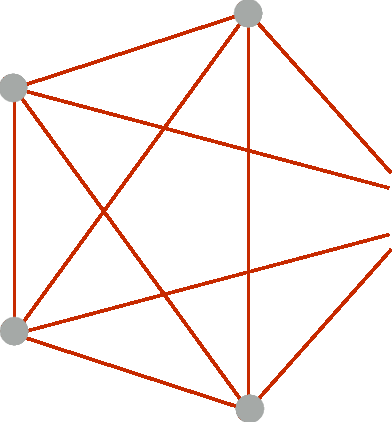}
			\caption{A brick-type pentagram network with 
			four open links.}
			\label{fig:pentagram_open}
\end{figure}

Here we propose an approach for constructing large Ising 
spin networks by utilizing smaller, pre-existing \emph{open} 
spin networks. This method involves transferring the 
quantum states in advance to reduce the number of qubits 
needed. To achieve this, we consider a ``brick" spin network 
with certain links left open, allowing for further 
connections to be made. For instance, in the case of the 
pentagram spin network, we can obtain the corresponding 
``brick" spin network by removing one of the nodes and 
leaving the four adjacent links open, as depicted in 
Fig. \ref{fig:pentagram_open}. By utilizing this technique, 
we can construct large-scale Ising spin networks more 
efficiently and with fewer qubits than previously possible.

It must be emphasized that our definition of an open spin 
network is a little different than some other used in the 
literature \cite{Bianchi:2023avf}. For example, in the 
case of an open single node, its state is equal to the 
state of a dipole with only one projection, i.e. contraction 
of four spin pairs with one intertwiner:
\begin{equation}
\iota_k^{m_{12}m_{13}m_{14}m_{15}}\alpha_{m_{12}m_2}
\alpha_{m_{13}m_3}\alpha_{m_{14}m_4}\alpha_{m_{15}m_5},
\end{equation}
where one spin from each pair (which is not contracted 
within the intertwiner) forms a boundary. So, in the 
definition applied here, boundary spins live at the ends 
of free links. Following the other definition it would 
be just that the intertwiner forming the boundary, \emph{i.e.}:
\begin{equation}
    \iota_k^{m_{12}m_{13}m_{14}m_{15}}.
\end{equation}
So, in other definition, the boundary is defined at the open 
node, which is not connected to other nodes.

\begin{figure}[ht!]
			\leavevmode
			\centering
			\Qcircuit @C=1em @R=1em {
				\lstick{\ket{0}} & \multigate{19}{U_\psi} & \multigate{3}{W^\dagger}&\qw&\qw\\
				\lstick{\ket{0}} & \ghost{U_\psi} & \ghost{W^\dagger}&\qw&\ket{0}&\bra{0}\\
				\lstick{\ket{0}} & \ghost{U_\psi} & \ghost{W^\dagger}&\qw&\ket{0}&\bra{0}\\
				\lstick{\ket{0}} & \ghost{U_\psi} & \ghost{W^\dagger}&\qw&\ket{0}&\bra{0}\\
				\lstick{\ket{0}} & \ghost{U_\psi} & \multigate{3}{W^\dagger}&\qw&\qw&\\
				\lstick{\ket{0}} & \ghost{U_\psi} & \ghost{W^\dagger} &\qw&\ket{0}&\bra{0}\\
				\lstick{\ket{0}} & \ghost{U_\psi} & \ghost{W^\dagger} &\qw&\ket{0}&\bra{0}\\
				\lstick{\ket{0}} & \ghost{U_\psi} & \ghost{W^\dagger} &\qw&\ket{0}&\bra{0}\\
				\lstick{\ket{0}} & \ghost{U_\psi} & \multigate{3}{W^\dagger}&\qw&\qw&\\
				\lstick{\ket{0}} & \ghost{U_\psi} & \ghost{W^\dagger} &\qw&\ket{0}&\bra{0}\\
				\lstick{\ket{0}} & \ghost{U_\psi} & \ghost{W^\dagger} &\qw&\ket{0}&\bra{0}\\
				\lstick{\ket{0}} & \ghost{U_\psi} & \ghost{W^\dagger} &\qw&\ket{0}&\bra{0}\\
				\lstick{\ket{0}} & \ghost{U_\psi} & \multigate{3}{W^\dagger}&\qw&\qw&\\
				\lstick{\ket{0}} & \ghost{U_\psi} & \ghost{W^\dagger} &\qw&\ket{0}&\bra{0}\\
				\lstick{\ket{0}} & \ghost{U_\psi} & \ghost{W^\dagger} &\qw&\ket{0}&\bra{0}\\
				\lstick{\ket{0}} & \ghost{U_\psi} & \ghost{W^\dagger} &\qw&\ket{0}&\bra{0}\\
				\lstick{\ket{0}} & \ghost{U_\psi} & \qw&\qw&\qw&\\
				\lstick{\ket{0}} & \ghost{U_\psi} & \qw &\qw&\qw\\
				\lstick{\ket{0}} & \ghost{U_\psi} & \qw &\qw&\qw\\
				\lstick{\ket{0}} & \ghost{U_\psi} & \qw &\qw&\qw\\
			}
			\caption{Partially projected state $|\psi\rangle=\bigotimes_{l\in\Gamma_5}\left|\mathcal{B},\frac{1}{2}\right\rangle$, i.e. pentagram with one open node}
			\label{fig:OpenPentagramCircuit}
		\end{figure}
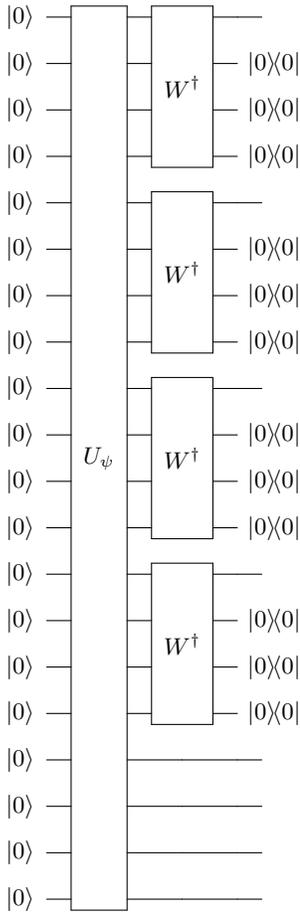

The quantum circuit corresponding to the brick-type (open) pentagram 
spin network is shown in Fig. \ref{fig:OpenPentagramCircuit}.
	
The pentagram spin network, with one removed node, 
can be accurately represented by an ansatz that 
utilizes only 8 qubits. This is a significant 
reduction compared to the unprojected case, which 
requires 20 qubits. In general, an $n$-node Ising 
spin network with one removed node can be represented 
by $n+3$ qubits. By connecting two networks, one 
with $n$ nodes and the other with $m$ nodes, 
we can obtain a network with $n+m$ nodes using 
$n+m+6$ qubits. In contrast, constructing the 
same network from individual singlet pairs would 
require $4(n+m)$ qubits. This represents a 
significant saving of quantum resources, with 
the amount saved being $3(n+m)-6$, particularly 
for larger networks. Importantly, this procedure 
can be iteratively applied to construct arbitrarily 
large networks by attaching the basis brick-type 
spin networks in succession. To illustrate this 
procedure, we provide an example using a 10-node 
spin network.

The use of partially projected states to construct 
a larger Ising spin network is possible because 
the projectors act on different nodes and therefore
commute. For the case of a 10-node Ising spin network 
we can prepare 20 links and then, apply 10 projections 
at once, or prepare, twice, 10 links for the pentagram, 
there apply four projections, and then apply the last 
two projections during the gluing:
\begin{equation}
\begin{split}
    \hat{P}_\Gamma^{10\otimes}&=\hat{\mathbb{I}}^{4\otimes}\otimes \hat{P}_\Gamma^{2\otimes}\otimes\hat{\mathbb{I}}^{4\otimes}\cdot\left[\left(\hat{P}_\Gamma^{4\otimes}\otimes\hat{\mathbb{I}}\right)\otimes\left(\hat{P}_\Gamma^{4\otimes}\otimes\hat{\mathbb{I}}\right)\right]\\ &=\left[\left(\hat{P}_\Gamma^{4\otimes}\otimes\hat{\mathbb{I}}\right)\otimes\left(\hat{P}_\Gamma^{4\otimes}\otimes\hat{\mathbb{I}}\right)\right]\cdot\hat{\mathbb{I}}^{4\otimes}\otimes \hat{P}_\Gamma^{2\otimes}\otimes\hat{\mathbb{I}}^{4\otimes}.
\end{split}
\end{equation}

\subsection{Gluing two pentagrams into a 10-node Ising spin network}

\begin{figure}
		\includegraphics[scale=0.7]{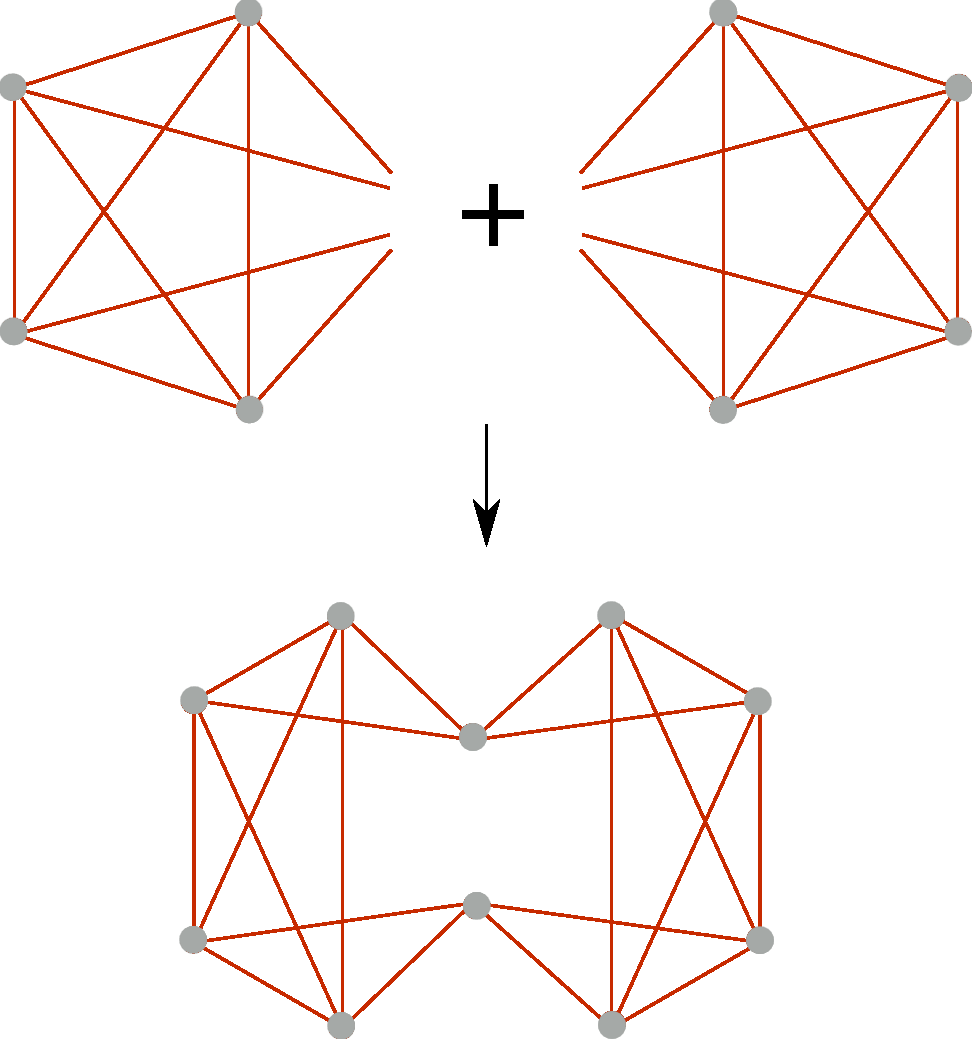}
		\caption{A 10-node spin network obtained by gluing two open pentagrams.}
		\label{fig:DekagramGluing}
	\end{figure}

Constructing a 10-node Ising spin network directly from 
the singlets at the links would require 40 logical qubits. 
This is already at the edge of the current capabilities 
of NISQ-type quantum technologies. However, by using the 
partial projection technique and gluing two brick-type 
pentagrams, we can generate the 10-node Ising spin 
network using only 16 qubits (8 qubits for each open 
pentagram). The procedure is illustrated in Fig. 
\ref{fig:DekagramGluing}. This significantly reduces the 
quantum resources required for constructing larger 
spin networks and makes it feasible with current quantum 
technologies.

\begin{figure}[ht!]
			\leavevmode
			\centering
			\Qcircuit @C=1em @R=1em {
				\lstick{\ket{0}} & \multigate{7}{\Gamma_5'}&\qw&\qw&\qw\\
				\lstick{\ket{0}} & \ghost{\Gamma_5'}&\qw&\qw&\qw\\
				\lstick{\ket{0}} & \ghost{\Gamma_5'}&\qw&\qw&\qw\\
				\lstick{\ket{0}} & \ghost{\Gamma_5'}&\qw&\qw&\qw\\
				\lstick{\ket{0}} & \ghost{\Gamma_5'}&\multigate{9}{W^\dagger}&\qw&\qw\\
				\lstick{\ket{0}} & \ghost{\Gamma_5'}&\ghost{W^\dagger}&\qw&\qw&\ket{0}&\bra{0}\\
				\lstick{\ket{0}} & \ghost{\Gamma_5'}&\qw&\multigate{9}{W^\dagger}&\qw\\
				\lstick{\ket{0}} & \ghost{\Gamma_5'}&\qw&\ghost{W^\dagger}&\qw&\ket{0}&\bra{0}\\
				\lstick{\ket{0}} & \multigate{7}{\Gamma_5'}&\qw&\qw&\qw\\
				\lstick{\ket{0}} & \ghost{\Gamma_5'}&\qw&\qw&\qw\\
				\lstick{\ket{0}} & \ghost{\Gamma_5'}&\qw&\qw&\qw\\
				\lstick{\ket{0}} & \ghost{\Gamma_5'}&\qw&\qw&\qw\\
				\lstick{\ket{0}} & \ghost{\Gamma_5'}&\ghost{W^\dagger}&\qw&\qw&\ket{0}&\bra{0}\\
				\lstick{\ket{0}} & \ghost{\Gamma_5'}&\ghost{W^\dagger}&\qw&\qw&\ket{0}&\bra{0}\\
				\lstick{\ket{0}} & \ghost{\Gamma_5'}&\qw&\ghost{W^\dagger}&\qw&\ket{0}&\bra{0}\\
				\lstick{\ket{0}} & \ghost{\Gamma_5'}&\qw&\ghost{W^\dagger}&\qw&\ket{0}&\bra{0}\\
			}
			\caption{Quantum circuit for gluing two open pentagrams.}
			\label{fig:dekagramcirc}
\end{figure}
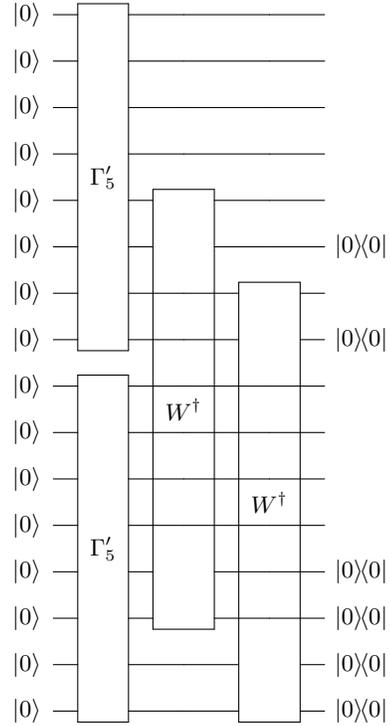

The quantum circuit associated with the gluing 
of two open pentagrams into the 10-node network 
is shown in Fig. \ref{fig:dekagramcirc}.

In the case of the 10-node Ising spin network, Fig.  
\ref{fig:DekagramGluing}, the projected state has the form:
\begin{equation}
    \hat{P}_{\Gamma}\left|\psi\right\rangle=\sum_{\iota_{k_i}}
    	c_{\iota_{k1}\iota_{k2}\iota_{k3}\iota_{k4}\iota_{k5}}\left|\iota_{k_1}\iota_{k_2}\iota_{k_3}\iota_{k_4}\iota_{k_5}\right\rangle,
\end{equation}
where 
\begin{equation}
\begin{split}
            c_{\iota_{k1}\iota_{k2}\iota_{k3}\iota_{k4}\iota_{k5}} &= \\
            &\iota_1^{m_{12}m_{13}m_{14}m_{15}}\iota_{2;m_{12}}^{\qquad m_{23}m_{24}m_{25}}\\
            &\iota_{3;m_{13}m_{23}}^{\qquad \qquad m_{34}m_{35}}\iota_{4;m_{14}m_{24}m_{34}}^{\qquad \qquad \qquad m_{45}}\\
            &\iota_{5;m_{15}m_{25}m_{610}m_{710}}\iota_{6;m_{35}m_{45}m_{810}m_{910}}\\
            &\iota_7^{m_{67}m_{68}m_{69}m_{610}}\iota_{8;m_{67}}^{\qquad m_{78}m_{79}m_{710}}\\
            &\iota_{9;m_{68}m_{78}}^{\qquad \qquad m_{89}m_{810}}\iota_{10;m_{69}m_{79}m_{89}}^{\qquad \qquad \qquad m_{910}}.\\
\end{split}
\label{eq:dekagram}
\end{equation}

The ansatz for the 10-node Ising spin network state is shown 
Fig. \ref{fig:s2d}, which has 5 layers and 10 qubits.
\begin{table}[h!]
    \centering
    \begin{tabular}{|c|c|c|c|c|c|c|c|c|c|c|}
        \hline
        2.52 & 2.15 &  & 0.11 &  & 0.78 &  & 2.86 &  &  1.75 &  \\
        \hline
        3.39 & 1.32 & 6.21 & 1.27 & 5.38 & 3.26 &  6.54 & 4.78 & -0.19 &  0.04 & 5.21 \\
        \hline
        1.28 & 4.34 & 4.11 & 6.43 & 1.81 & 0.24 &  1.59 & 2.04 &  3.74 &  2.02 & 4.37 \\
        \hline
        4.76 & 2.00 & 6.28 & 3.14 & 1.20 & 0.83 & -0.13 & 3.88 &  2.50 &  3.13 & 0.60 \\
        \hline
        5.65 & 4.71 & 3.14 & 3.14 & 5.34 & 3.14 &  4.72 & 1.57 &  6.28 &  1.57 & 1.78 \\
        \hline
        3.14 & 3.14 & 6.28 & 0.00 & 4.62 & 4.80 &  3.21 & 1.57 &  6.28 &  0.97 & 4.92 \\
        \hline
        0.27 & 5.03 & 5.92 & 3.80 & 5.89 & 1.71 &  0.57 & 0.73 & -0.02 &  2.21 & 4.57 \\
        \hline
        5.64 & 5.07 & 1.78 & 1.68 & 1.39 & 5.72 &  5.20 & 3.37 &  5.95 & -0.36 & 6.70 \\
        \hline
        -0.31 & 5.73 & 2.36 & 5.08 & 3.53 & 0.81 &  0.84 & 3.07 &  1.44 &  5.06 & 0.33 \\
        \hline
        4.08 & 3.35 &  & 1.32 &  & 2.12 &  & 0.68 &  &  4.56 & \\
        \hline
    \end{tabular}
    \caption{The parameters found for the 10-node network ansatz. 
    The structure of table corresponds to structure of ansatz 
    Fig.~\ref{fig:s2d}, with 5 layers and 10 qubits.}
    \label{tab:pentagram_params}
\end{table}

The fidelity between the states obtained from the quantum 
algorithm running on the classical simulator without noise 
and the states obtained from Eq. \ref{eq:dekagram} is 
approximately 0.9975. It is reasonable to assume that 
the fidelity is essentially 1, up to numerical errors.

\subsection{Gluing single nodes into an arbitrary network}

\begin{figure}[ht!]
 	\centering
 		\includegraphics[scale=0.7]{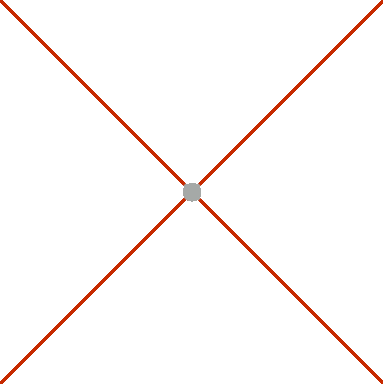}
 		\caption{One node with four open links.}
 		\label{fig:opennode}
 	\end{figure}

Another possibility in construction of the Ising spin 
networks is to prepare an ansatz circuit of a single 
node with four open links, depicted Fig. \ref{fig:opennode}.

The corresponding state can be expressed as follows:
\begin{equation}
\begin{split}
    \left|\Gamma_2'\right\rangle= \sum_{\iota_k} &\iota_k^{m_{12}m_{13}m_{14}m_{15}}\\&\alpha_{m_{12}m_2}\alpha_{m_{13}m_3}\alpha_{m_{14}m_4}\alpha_{m_{15}m_5}\\
     &\left|\iota_k\right\rangle\left|\frac{1}{2},m_2\right\rangle\left|\frac{1}{2},m_3\right\rangle\left|\frac{1}{2},m_4\right\rangle\left|\frac{1}{2},m_5\right\rangle.\\
\end{split}
\end{equation}

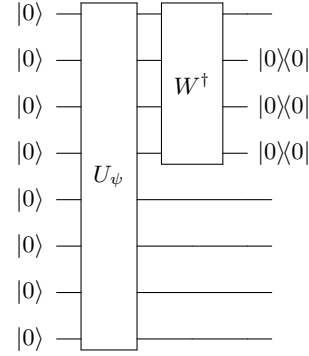
\begin{figure}[ht!]
 			\leavevmode
 			\centering
 			\Qcircuit @C=1em @R=1em {
 				\lstick{\ket{0}} & \multigate{7}{U_\psi} & \multigate{3}{W^\dagger}&\qw&\qw\\
 				\lstick{\ket{0}} & \ghost{U_\psi} & \ghost{W^\dagger}&\qw&\ket{0}&\bra{0}\\
 				\lstick{\ket{0}} & \ghost{U_\psi} & \ghost{W^\dagger}&\qw&\ket{0}&\bra{0}\\
 				\lstick{\ket{0}} & \ghost{U_\psi} & \ghost{W^\dagger}&\qw&\ket{0}&\bra{0}\\
 				\lstick{\ket{0}} & \ghost{U_\psi} & \qw&\qw&\qw&\\
 				\lstick{\ket{0}} & \ghost{U_\psi} & \qw &\qw&\qw\\
 				\lstick{\ket{0}} & \ghost{U_\psi} & \qw &\qw&\qw\\
 				\lstick{\ket{0}} & \ghost{U_\psi} & \qw &\qw&\qw\\
 			}
 			\caption{Quantum circuit of a single node with four open links.}
 			\label{fig:opennode_circ}
 		\end{figure}

Based on the 8-qubit circuit shown in Fig. \ref{fig:opennode_circ} 
the 5-qubit ansatz circuit can be constructed, which saves 3 qubits 
with respect to the original approach. 

Gluing nodes together can result in an arbitrary Ising spin 
network state. However, it may not always be the most optimal 
way to construct the network. For instance, gluing two open 
nodes together to form a dipole network would require 10 qubits 
instead of the 8 qubits needed for the basic method, which 
does not provide an advantage. However, in cases where a 
larger network has already been constructed using the open 
pentagram networks, adding a single node may prove to be advantageous.

\begin{figure}[ht!]
    \centering
 	\includegraphics[scale=0.7]{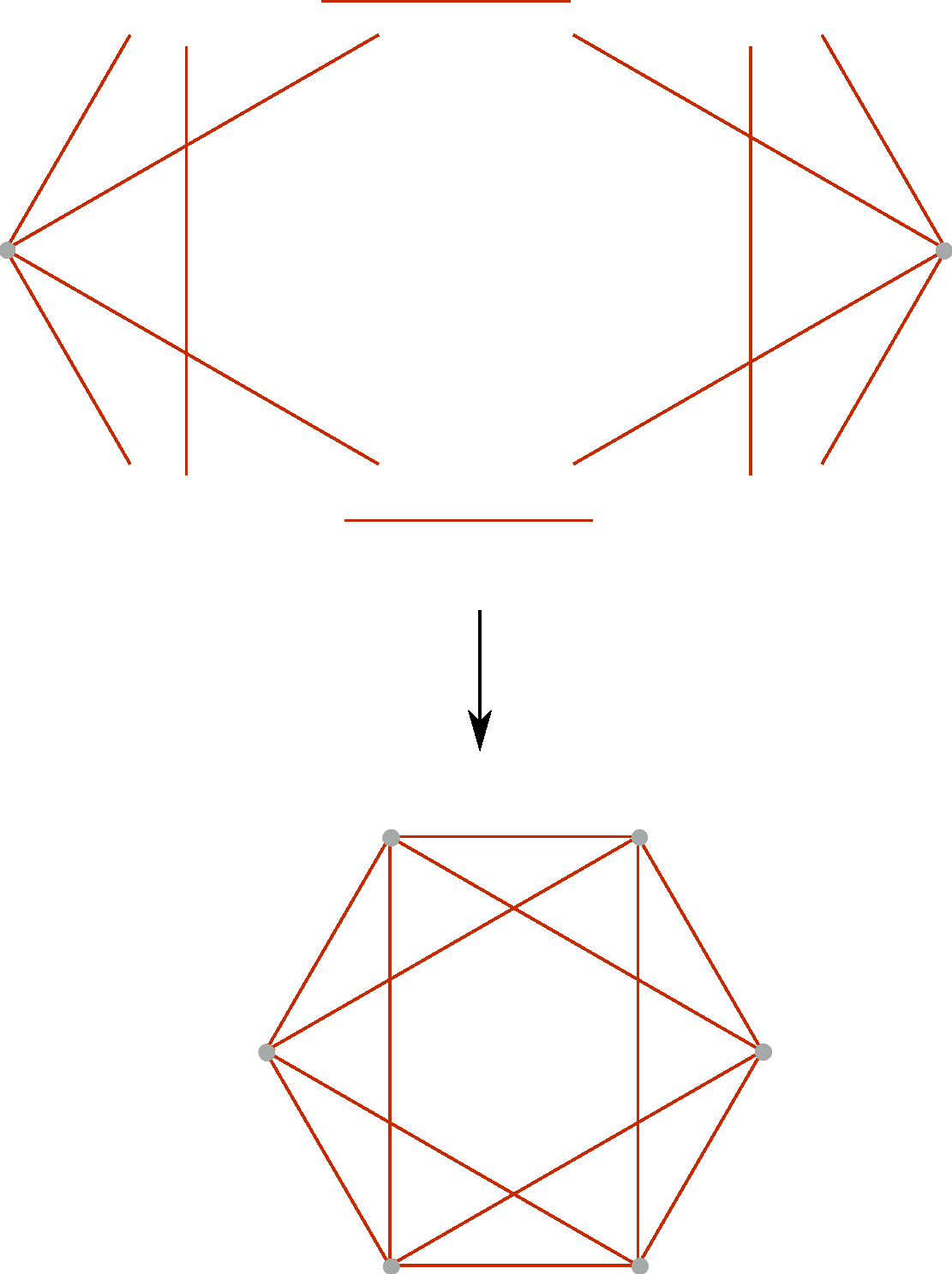}
 	\caption{Hexagram network obtained from two nodes each with four free links and from four additional links.}
 	\label{fig:hexagram}
\end{figure}

To illustrate this, let us consider the hexagram network, 
which consists of 6 nodes. The construction scheme of the 
network is depicted in Fig. \ref{fig:hexagram}, where two 
nodes and four free links are involved.

\begin{figure*}[ht!]
 	\leavevmode
 	\centering
 	\Qcircuit @C=1em @R=1em {
     	\lstick{\ket{0}} & \multigate{4}{\Gamma_2'} & \qw\ & \qw & \qw & \qw & \qw & \qw & \qw & \qw & \qw  & \qw \\
     	\lstick{\ket{0}} & \ghost{\Gamma_2'} & \qw & \qw & \qw & \qw & \qw & \multigate{10}{W^\dagger} & \qw & \qw & \qw & \qw \\
     	\lstick{\ket{0}} & \ghost{\Gamma_2'} & \qw & \qw & \qw & \qw & \qw & \qw & \multigate{13}{W^\dagger} & \qw & \qw & \qw \\
     	\lstick{\ket{0}} & \ghost{\Gamma_2'} & \qw & \qw & \qw & \qw & \qw & \qw & \qw & \multigate{13}{W^\dagger} & \qw & \qw \\
     	\lstick{\ket{0}} & \ghost{\Gamma_2'} & \qw & \qw & \qw & \qw & \qw & \qw & \qw & \qw & \multigate{13}{W^\dagger} & \qw \\
     	\lstick{\ket{0}} & \multigate{4}{\Gamma_2'} & \qw & \qw & \qw & \qw & \qw & \qw & \qw & \qw & \qw & \qw \\
     	\lstick{\ket{0}} & \ghost{\Gamma_2'} & \qw & \qw & \qw & \qw & \qw & \qw & \ghost{W^\dagger} & \qw & \qw  & \qw  & \ket{0}&\bra{0}\\
     	\lstick{\ket{0}} & \ghost{\Gamma_2'} & \qw & \qw & \qw & \qw & \qw & \ghost{W^\dagger} & \qw & \qw & \qw  & \qw  & \ket{0}&\bra{0}\\
     	\lstick{\ket{0}} & \ghost{\Gamma_2'} & \qw & \qw & \qw & \qw & \qw & \qw & \qw & \qw & \ghost{W^\dagger}  & \qw  & \ket{0}&\bra{0}\\
     	\lstick{\ket{0}} & \ghost{\Gamma_2'} & \qw & \qw & \qw & \qw & \qw & \qw & \qw & \ghost{W^\dagger} & \qw  & \qw  & \ket{0}&\bra{0}\\
     	\lstick{\ket{0}} & \gate{X} & \gate{H} & \ctrl{4} & \qw & \qw & \qw & \ghost{W^\dagger} & \qw & \qw & \qw  & \qw  & \ket{0}&\bra{0}\\
     	\lstick{\ket{0}} & \gate{X} & \gate{H} & \qw & \ctrl{4} & \qw & \qw & \ghost{W^\dagger} & \qw & \qw & \qw  & \qw  & \ket{0}&\bra{0}\\
     	\lstick{\ket{0}} & \gate{X} & \gate{H} & \qw & \qw & \ctrl{4} & \qw & \qw & \ghost{W^\dagger} & \qw & \qw  & \qw  & \ket{0}&\bra{0}\\
     	\lstick{\ket{0}} & \gate{X} & \gate{H} & \qw & \qw & \qw & \ctrl{4} & \qw & \qw & \ghost{W^\dagger} & \qw  & \qw  & \ket{0}&\bra{0}\\
     	\lstick{\ket{0}} & \gate{X} & \qw & \targ & \qw & \qw & \qw & \qw & \qw & \qw & \ghost{W^\dagger} & \qw  & \ket{0}&\bra{0}\\
     	\lstick{\ket{0}} & \gate{X} & \qw & \qw & \targ & \qw & \qw & \qw & \ghost{W^\dagger} & \qw & \qw & \qw  & \ket{0}&\bra{0}\\
     	\lstick{\ket{0}} & \gate{X} & \qw & \qw & \qw & \targ & \qw & \qw & \qw & \ghost{W^\dagger} & \qw & \qw  & \ket{0}&\bra{0}\\
     	\lstick{\ket{0}} & \gate{X} & \qw & \qw & \qw & \qw & \targ & \qw & \qw & \qw & \ghost{W^\dagger} & \qw  & \ket{0}&\bra{0}\\
 	}
 	\caption{Hexagram network obtained from two nodes, each with four free links 
  ($\Gamma_2'$) and from four additional links.}
 	\label{fig:hexagram_circ}
\end{figure*}
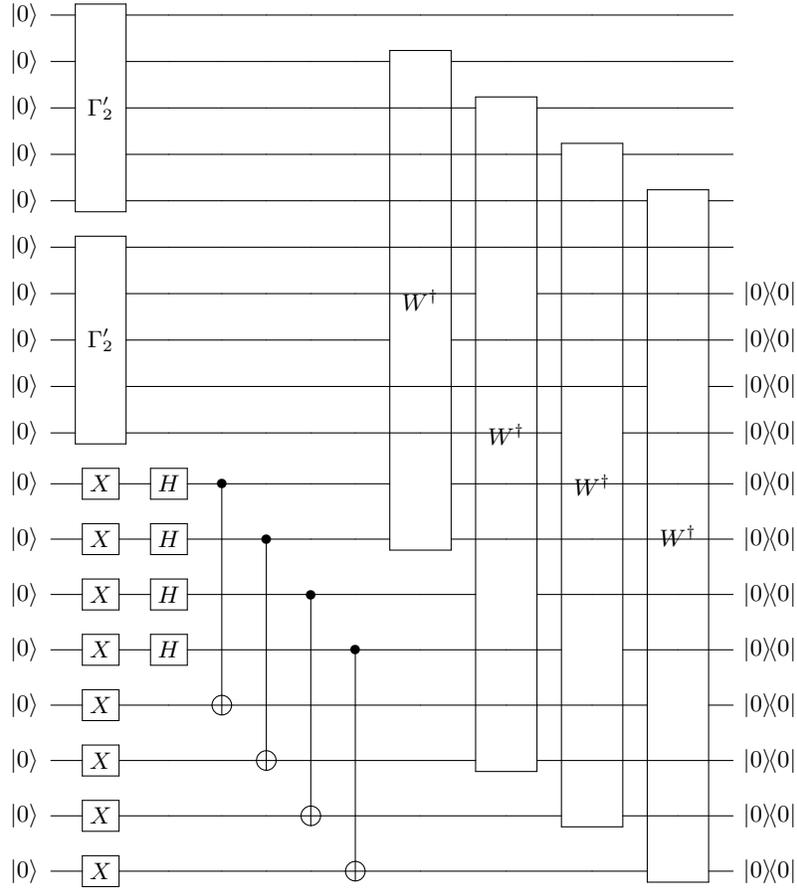

The construction utilizes $2 \cdot 5+4 \cdot 2=18$ qubits, which 
is a bit smaller than for the basic method, involving $10 \cdot 2=20$ 
qubits. The corresponding 18-qubit quantum circuit is shown in Fig. 
\ref{fig:hexagram_circ}.
 
It is important to note that to minimize the quantum resources 
needed to construct an arbitrary Ising spin network, one should 
consider the optimal combination of pre-existing open spin 
networks and free links. This requires careful consideration of 
the available resources and the desired network topology. 
However, determining the optimal sequence of gluing is a separate 
task that must be addressed individually, and depends on the 
specific details of the network being constructed. Finding the 
optimal strategy for constructing an Ising spin network 
with minimal quantum resources is an ongoing research challenge.

\section{Results}
\label{Sec:Results}

\subsection{Dipole}

\begin{figure}
	\includegraphics[scale=0.5]{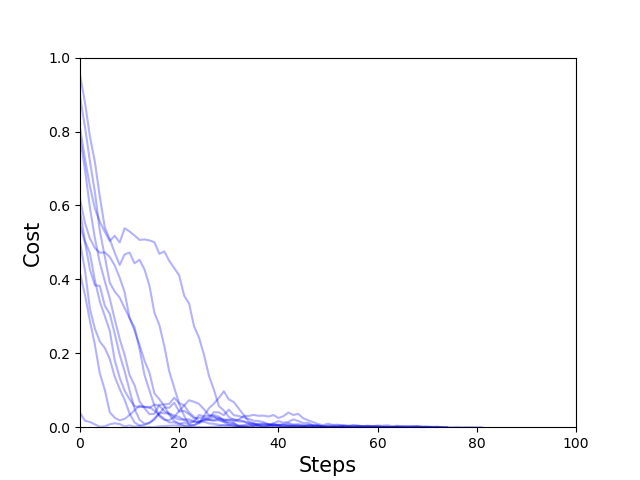}
	\caption{The cost function for the dipole transfer with statistical noise for ten independent simulations with randomly initialized parameters.}
	\label{fig:dipole_cost}
\end{figure}

We employed the circuit illustrated in Fig. \ref{fig:dipole_transfer} 
to optimize a quantum device simulator, which was subject to statistical 
noise due to a finite number of shots, but not quantum noise. We used 
the PennyLane library for the simulations, which were run on an eight-qubit 
quantum device with 20,000 shots per circuit execution. To compute 
the gradient of the cost function, we utilized the parameter shift rule, 
while the classical Adam optimizer with a step size of 0.1 was chosen as 
the optimizer. Each optimization was halted after achieving convergence 
below $10^{-6}$ or after 100 steps. Fig. \ref{fig:dipole_cost} displays 
the history of the cost function for ten independent simulations with 
randomly initialized parameters.

\begin{figure}
	\includegraphics[scale=0.5]{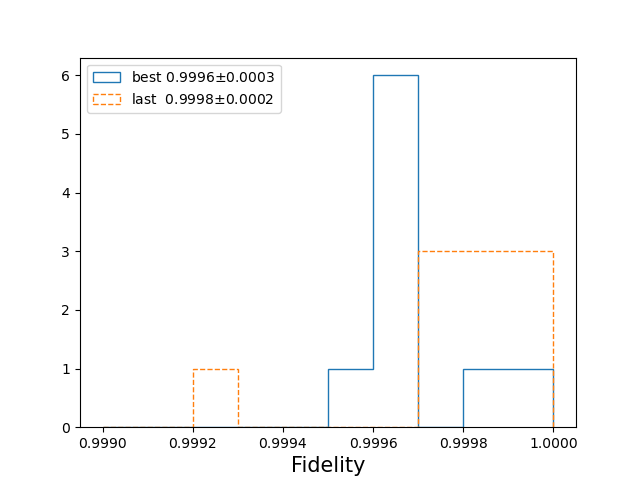}
	\caption{Histogram of fidelities of dipole states obtained, at the last step (orange, dashed line), or at the step we the minimal cost value (blue, solid line).}
	\label{fig:dipole_hist}
\end{figure}

After each minimization, we obtained the precise state described 
by the optimized parameter ansatz and calculated its fidelity 
with the expected theoretical state. Fig. \ref{fig:dipole_hist} 
presents the mean fidelities and corresponding standard deviations 
obtained from ten independent simulations of the dipole.

\subsection{Pentagram}

\begin{figure}
	\includegraphics[scale=0.5]{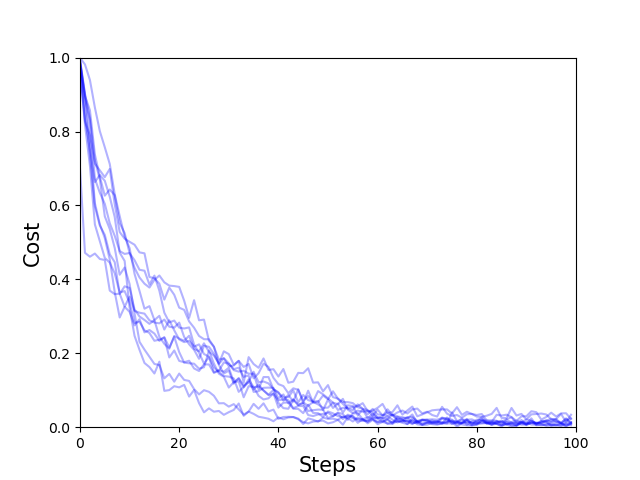}
	\caption{The cost function for the pentagram transfer with 
 statistical noise for ten independent simulations with randomly 
 initialized parameters.}
	\label{fig:pentagram_cost}
\end{figure}

Here, we utilized the pentagram construction scheme 
illustrated in Fig. \ref{fig:pentagram}. The state 
corresponding to the open node utilized in the 
procedure was obtained using an exact simulator 
devoid of statistical noise. The simulations were 
executed on a 17-qubit quantum device simulator 
with 200,000 shots per circuit execution. To 
calculate the gradient of the cost function, 
we used the parameter shift rule, and we chose 
the classical Adam optimizer with a step size 
of 0.1 as the optimizer. Each minimization was 
terminated after achieving convergence below 
$10^{-6}$ or after 100 steps. Fig. \ref{fig:pentagram_cost} 
displays the cost function's optimization 
history for ten independent simulations with 
randomly initialized parameters.

\begin{figure}[ht!]
    \centering
 	\includegraphics[scale=0.7]{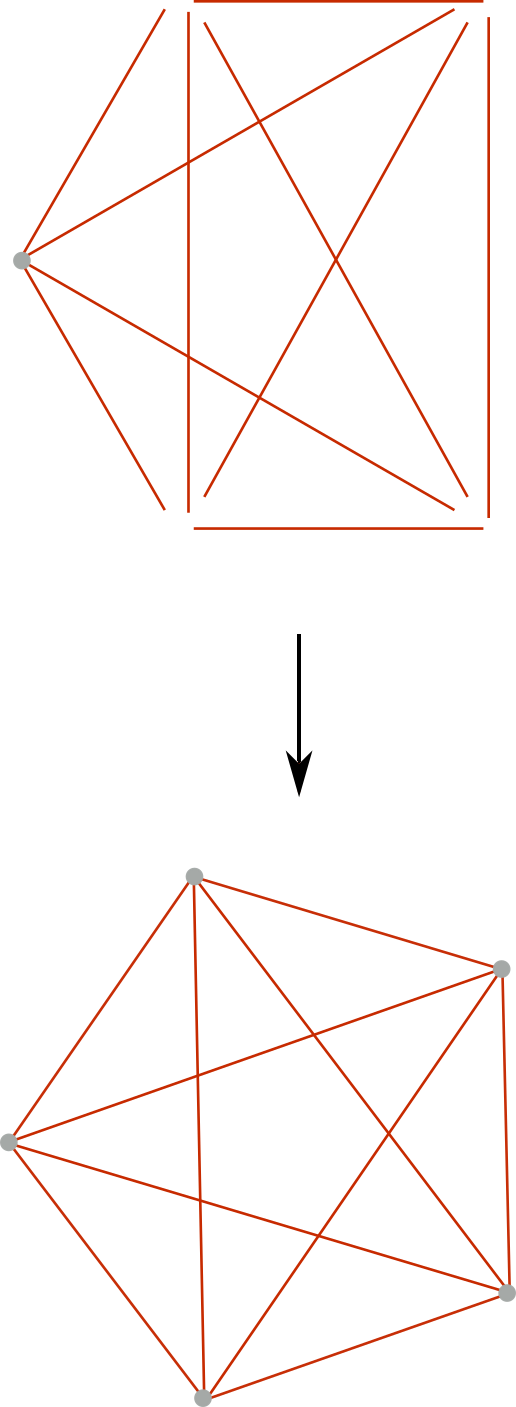}
 	\caption{Pentagram network obtained from a node with four free links and from six additional links.}
 	\label{fig:pentagram}
 \end{figure}

\begin{figure}
	\includegraphics[scale=0.5]{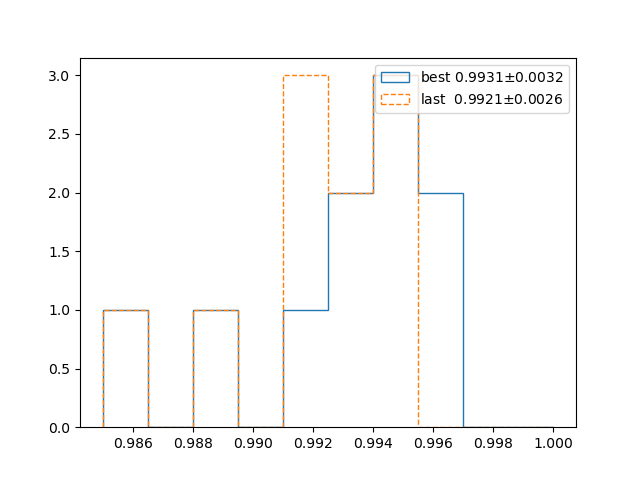}
	\caption{Histogram of fidelities of pentagram states obtained, at the last step (orange, dashed line), or at the step we the minimal cost value (blue, solid line).}
	\label{fig:pentagram_hist}
\end{figure}

Fig. \ref{fig:pentagram_hist} shows the mean fidelities and standard 
deviations computed from ten independent simulations of the dipole problem.

\subsection{Hexagram}

\begin{figure}
	\includegraphics[scale=0.5]{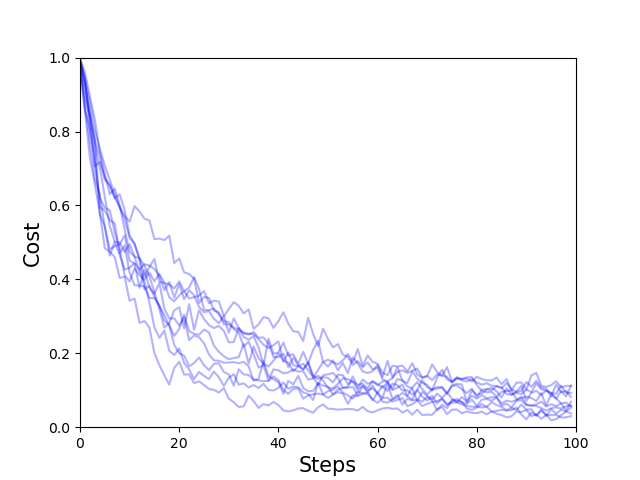}
	\caption{The cost function for the hexagram transfer with 
 statistical noise for ten independent simulations with randomly 
 initialized parameters.}
	\label{fig:hexagram_cost}
\end{figure}

For this problem, we adopted the scheme depicted in 
Fig. \ref{fig:hexagram}. The state corresponding to 
the open node, which we employed in this procedure, 
was obtained using an exact simulator without statistical 
noise. The simulations were executed on a quantum 
device simulator with 18 qubits, and each circuit 
execution comprised 200,000 shots. To compute the 
cost function's gradient, we employed the parameter 
shift rule, while we chose the classical Adam 
optimizer with a step size of 0.1 as the optimizer. 
Each minimization was halted after achieving 
convergence below $10^{-6}$ or after 100 steps. 
Fig. \ref{fig:hexagram_cost} presents the history 
of the cost function's optimization for ten 
independent simulations with randomly initialized parameters.

\begin{figure}
	\includegraphics[scale=0.5]{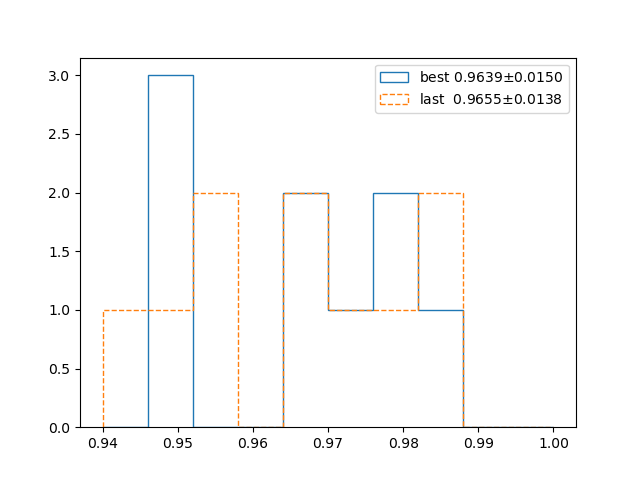}
	\caption{Histogram of fidelities of hexagram states obtained, at the last step (orange, dashed line), or at the step we the minimal cost value (blue, solid line). In this case both lines overlap.}
	\label{fig:hexagram_hist}
\end{figure}

Fig. \ref{fig:hexagram_hist} illustrates the mean fidelities and corresponding standard deviations obtained from ten independent simulations of the hexagram problem.

\subsection{A 10-node graph}

\begin{figure}
	\includegraphics[scale=0.5]{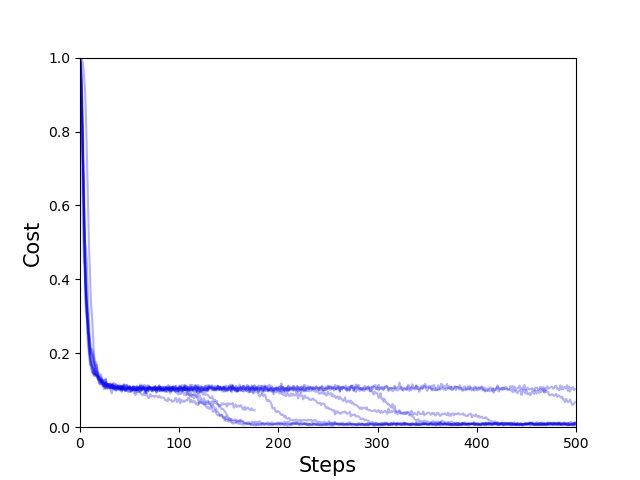}
	\caption{The cost function for 10-node Ising spin network state
 transfer with statistical noise for ten independent simulations 
 with randomly initialized parameters.}
	\label{fig:dekagram_cost}
\end{figure}

For this problem, we employed the scheme shown 
in Fig. \ref{fig:DekagramGluing}. The state 
corresponding to the open pentagram, which 
we used in this procedure, was obtained using 
an exact simulator without statistical noise. 
The simulations were executed on a quantum device
simulator with 16 qubits, and each circuit 
execution comprised 20,000 shots. To compute 
the cost function's gradient, we employed the 
parameter shift rule, and we used the classical 
Adam optimizer with a step size of 0.1 as the 
optimizer. Each minimization was halted after 
achieving convergence below $10^{-6}$ or 
after 500 steps. Fig. \ref{fig:dekagram_cost} 
presents the history of the cost function's 
optimization for ten independent simulations 
with randomly initialized parameters.

\begin{figure}
	\includegraphics[scale=0.5]{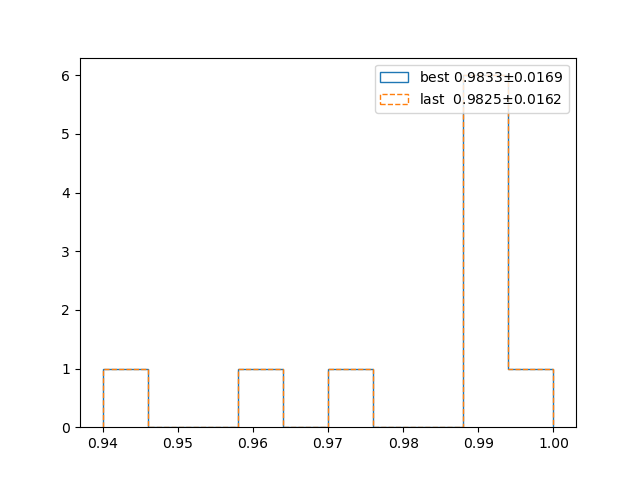}
	\caption{Histogram of fidelities of 10-node spin network states obtained
	at the last step (orange, dashed line), or at the step we the minimal 
	cost value (blue, solid line). In this case both lines overlap.}
	\label{fig:dekagram_hist}
\end{figure}

Fig. \ref{fig:dekagram_hist} displays the mean fidelities 
and corresponding standard deviations obtained from ten 
independent simulations of the 10-node problem.

\section{Summary}
\label{Sec:Summary}

This article presents an optimized procedure for constructing 
quantum circuits that generate Ising spin network states. 
These states are characterized by four-valent nodes with 
qubit degrees of freedom attached to them. Although other 
types of spin networks are more general, Ising spin network 
states are sufficient to model some important properties 
of quantum space, particularly with respect to the structure 
of quantum correlations, entanglement entropy, and the 
quantum thermodynamic limit.

The investigation was primarily motivated by the potential 
application of quantum computers in simulating complex spin 
network states, which are currently beyond the capabilities 
of even the most powerful classical supercomputers. However, 
a key challenge in this field is the need to better understand 
the computational complexity of the relevant quantum amplitudes 
associated with spin network states. To determine the exact 
computational cost of such simulations, we must investigate 
the quantum circuit representation of these networks. By 
doing so, we can establish an upper constraint on the quantum 
complexity of these simulations and potentially even determine 
their exact value in certain cases.

An inefficient method for generating Ising spin network states 
involves using $4n$ qubits to project link states and create 
a spin network with $n$ nodes. In contrast, the method we 
introduce significantly reduces the number of required qubits, 
leading to practical benefits. Additionally, the introduced 
projection operator enables the transfer of the final state 
onto an $n$-qubit ansatz circuit, allowing for further 
operations without the need for redundant qubits. This 
ansatz has numerous applications, including analyzing 
the quantum complexity of quantum-gravitational transition 
amplitudes and studying the fate of quantum fluctuations 
in geometric quantities and the scaling of quantum entropy. 
Our future studies will delve deeper into these topics.

In addition to the issues discussed so far, an important 
future direction of investigation is the study of partially 
projected states within the developed framework. In these 
states, the degrees of freedom at open links can be 
interpreted as boundary values, while the remaining 
network corresponds to the bulk. This approach has 
promising implications for understanding the possible 
holographic nature of gravitational interactions, 
and can help address related problems.

Generalizations of the spin network state construction 
presented here to higher valence cases, beyond 
the Ising approximation, are also worth exploring. 
Additionally, future improvements in the methods of 
generating these states may benefit from utilizing 
the tensor network approach \cite{Han:2016xmb,Colafranceschi:2021acz}.

\section*{ACKNOWLEDGMENTS}

The research has been supported by the Sonata Bis
Grant No. DEC-2017/26/E/ST2/00763 of the National
Science Centre Poland. This research was funded by 
the Priority Research Area Digiworld under the 
program Excellence Initiative – Research University 
at the Jagiellonian University in Krak\'ow.
\\

\section*{Appendix}

\subsection*{Decomposition of $\sqrt{SWAP}$ gate}

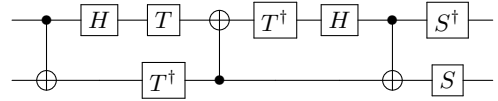
\begin{figure}[ht!]
 			\leavevmode
 			\centering
 			\Qcircuit @C=1em @R=1em {
 				& \ctrl{1} & \gate{H} & \gate{T} & \targ & \gate{T^\dagger} & \gate{H} & \ctrl{1} & \gate{S^\dagger} & \qw\\
 				& \targ & \qw & \gate{T^\dagger} & \ctrl{-1} & \qw & \qw & \targ & \gate{S} & \qw\\
 			}
 			\caption{Decomposition of $\sqrt{SWAP}$ gate.}
 			\label{fig:sqrtSWAP}
\end{figure}

In Fig. \ref{fig:sqrtSWAP} we present an explicit form 
of a circuit for the $\sqrt{SWAP}$ gate, expressed in 
terms of the elementary single- and two-qubit gates. 

\subsection*{Simplified circuit for the operator $\hat{W}$}

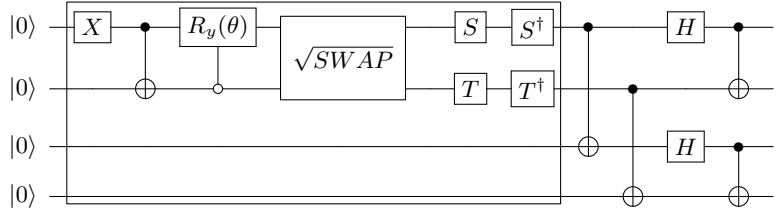
\begin{figure*}[ht!]
    \leavevmode
    \centering
    \Qcircuit @C=1em @R=1em {
        \lstick{\ket{0}}  & \gate{X} & \ctrl{1} & \gate{R_y(\theta)} & \multigate{1}{\sqrt{SWAP}} &  \qw & \gate{S} & \gate{S^\dagger} & \ctrl{2} & \qw & \gate{H} & \ctrl{1} & \qw\\
        \lstick{\ket{0}} & \qw & \targ & \ctrlo{-1} & \ghost{\sqrt{SWAP}} & \qw & \gate{T} & \gate{T^\dagger} & \qw & \ctrl{2} & \qw & \targ & \qw\\
        \lstick{\ket{0}} & \qw & \qw & \qw & \qw & \qw & \qw & \qw & \targ & \qw & \gate{H} & \ctrl{1} & \qw\\
        \lstick{\ket{0}} & \qw & \qw & \qw & \qw & \qw & \qw & \qw & \qw & \targ &\qw & \targ &\qw
        \gategroup{1}{2}{4}{8}{.6em}{-}
    }
    \caption{Quantum circuit for the operator $\hat{W}$ with the indication 
    of the part being a subject of variational approximation.}
    \label{fig:Wvar}
\end{figure*}

Due to its complexity, applying operator $\hat{W}$ (see Fig. 
\ref{fig:Wacts}) on current noisy quantum processors may 
lead to numerous errors. To overcome this issue, a simpler 
circuit that performs $\hat{W}$ can be found using variational 
methods \cite{khatri2019quantum}. Fig. \ref{fig:Wvar} illustrates 
the most intricate part of the circuit that needs to be simplified.

\begin{figure}[ht!]
    \leavevmode
    \centering
    \Qcircuit @C=1em @R=1em {
        \lstick{\ket{0}}  & \gate{R_y\left(\theta\right)} & \ctrl{1} & \gate{R_y\left(\theta\right)} & \ctrl{1} &  \gate{R_y\left(\theta\right)} &  \ctrl{2} & \qw & \gate{H} & \ctrl{1} & \qw\\
        \lstick{\ket{0}} & \gate{R_y\left(\theta\right)} & \ctrl{-1} & \gate{R_y\left(\theta\right)} & \ctrl{-1} & \gate{R_y\left(\theta\right)} & \qw & \ctrl{2} & \qw & \targ & \qw\\
        \lstick{\ket{0}}  & \qw & \qw & \qw & \qw & \qw & \targ & \qw & \gate{H} & \ctrl{1} & \qw\\
        \lstick{\ket{0}} & \qw & \qw & \qw & \qw & \qw & \qw & \targ &\qw & \targ &\qw
        \gategroup{1}{2}{4}{6}{.6em}{-}
    }
    \label{fig:Wvart}
    \caption{Quantum circuit for the operator $\hat{W}$ with the employed 
    variational ansatz circuit.}
\end{figure}
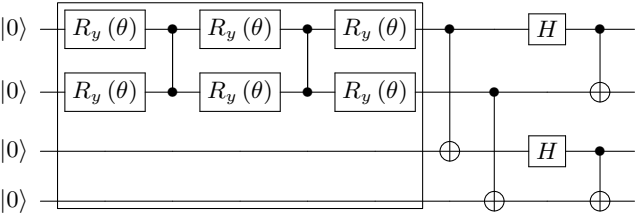

To simplify the circuit and reduce the likelihood of errors on 
noisy quantum processors, we can substitute the intricate section 
with an ansatz circuit. By minimizing the cost function, we can 
determine the parameters that allow the new circuit to perform 
similarly to the operator $\hat{W}$, although it does not need 
to be an identical operator. The primary goal is to rotate 
one-qubit states into the intertwiner subspace of four qubits. 
The resulting circuit has the form depicted in Fig. \ref{fig:Wvart}.

\end{document}